\documentclass[aip, amsmath,amssymb, reprint]{revtex4-1}

\usepackage{booktabs}
\usepackage{graphicx}% Include figure files
\usepackage{xcolor}
\usepackage{dcolumn}% Align table columns on decimal point
\usepackage{bm}% bold math
\usepackage{natbib}
\usepackage{multirow} 
\setcitestyle{super}

\usepackage{tikz}
\usetikzlibrary{decorations.markings}
\usetikzlibrary{calc}
\usetikzlibrary{arrows.meta, positioning}
\usepackage{array}
\usepackage{braket}
\usepackage{enumitem}
\setlist{leftmargin=4mm}
\usepackage{afterpage}
\usepackage{placeins}
\usepackage[normalem]{ulem}
\newcolumntype{L}[1]{>{\raggedright\let\newline\\\arraybackslash\hspace{0pt}}m{#1}}
\newcolumntype{C}[1]{>{\centering\let\newline\\\arraybackslash\hspace{0pt}}m{#1}}
\newcolumntype{R}[1]{>{\raggedleft\let\newline\\\arraybackslash\hspace{0pt}}m{#1}}

\begin{document}

\preprint{AIP/123-QED}

\title{Elucidating Many-Body Effects in Molecular Core Spectra through Real-Time Approaches: Efficient Classical Approximations and a Quantum Perspective}

\author{Vibin Abraham}
\email{vibin.abraham@pnnl.gov}
\affiliation{Physical and Computational Science Directorate, Pacific Northwest National Laboratory, Richland, Washington 99354, USA}

\author{Priyabrata Senapati}
\affiliation{Physical and Computational Science Directorate, Pacific Northwest National Laboratory, Richland, Washington 99354, USA}
\affiliation{Department of Computer Science, Kent State University, Kent, Ohio 44240, USA} 

\author{Himadri Pathak}
\affiliation{Quantum Mathematical Science Team, Division of Applied Mathematical Science, RIKEN Center for Interdisciplinary Theoretical and Mathematical Sciences (iTHEMS), 2-1 Hirosawa Wako, Saitama 351-0198, Japan}
\affiliation{Computational Molecular Science Research Team, RIKEN Center for Computational Science (R-CCS), 7-1-26 Minatojima-minami-machi, Chuo-ku, Kobe, Hyogo 650-0047, Japan}

\author{Bo Peng}
\email{peng398@pnnl.gov}

\affiliation{Physical and Computational Science Directorate, Pacific Northwest National Laboratory, Richland, Washington 99354, USA} 

\date{\today}
\begin{abstract}
Accurately resolving many-body satellite features in molecular core-level spectra requires theoretical approaches that capture electron correlation both efficiently and systematically. The recently developed time-dependent double coupled-cluster (TD-dCC) ansatz achieves this by combining correlation effects from the $N$- and $(N{-}1)$-electron sectors, but its exact formulation remains computationally demanding. Here we introduce a hierarchy of cost-effective approximate TD-dCC ans\"{a}tzes derived from truncated Baker–Campbell–Hausdorff (BCH) expansions, which preserve a single-similarity-transformation structure while retaining the essential correlation diagrams responsible for satellite formation. We further develop a detailed component analysis that isolates hole-mediated excitation pathways---correlated processes arising from the coupling between ground-state and ionized-state amplitudes---and use it to interpret quasiparticle and satellite features across the hierarchy. Applications to the single-impurity Anderson model and molecular systems (H$_2$O and CH$_4$) demonstrate that the approximate TD-dCC methods closely and efficiently reproduce exact many-body spectral features and quasiparticle weights. In parallel, we construct a fault-tolerant quantum signal processing algorithm for the core-hole Green’s function, providing a scalable quantum route for simulating correlated core-level dynamics. Together, these developments establish complementary classical and quantum methodologies for quantitative, many-body-accurate core spectroscopy.
\end{abstract}

\maketitle

%%%%%%%%%%%%%%%%%%%%%%%%%%%%

\section{Introduction}\label{sec:intro}

X-ray spectroscopies provide a powerful window into the electronic structure of molecules and materials, offering direct access to quasiparticle energies and many-body satellite features that reveal electron correlation and collective excitation effects~\cite{butanovs2021nuclear,greczynski2020x}. Recent developments in ultrafast attosecond X-ray techniques have further extended these methods beyond static characterization, enabling real-time observation of electronic responses to ionization, charge migration, excitation dynamics, \textcolor{black}{and} insulator-metal transitions~\cite{mayer2022following,schwickert2022electronic,ridente2023femtosecond,mazzaro2025operando,gray2016correlation}. For example, attosecond X-ray pump–probe experiments are capable of resolving correlated electron dynamics in $<$1 fs, providing direct observation of the ultrafast evolution of ionized states in condensed systems such as liquid water~\cite{li2024attosecond,guo2024experimental}. To connect these experimental measurements to the underlying electronic structure, there has been growing effort in developing first-principles computational methods for quantitatively interpreting X-ray spectroscopies~\cite{rehr2009ab,besley2020density,prendergast2006,wernet2015orbital,coriani2015communication,simons2023transition,ranga2021core,mazin2023core, Golze2020,van2018assessing,ehara2006c1s,ahmed2025core}.

The central theoretical challenge lies in capturing the complex many-body interactions that govern both quasiparticle and satellite features. Satellite excitations arise from electron correlation effects that fundamentally exceed the scope of mean-field approaches such as Hartree–Fock (HF) or density functional theory (DFT). This limitation has driven the development of numerous advanced theoretical frameworks that not only predict ionization spectra with high accuracy, but also elucidate the microscopic mechanisms underlying quantum many-body phenomena~\cite{cederbaum1977complete,mejuto2021multi,bagus2022origin,ghiasi2019charge}. Moreover, many X-ray spectroscopies involve core-hole–mediated transitions~\cite{woicik2020core,unger2017observation,Esther2025}, whose transient, highly correlated character makes them difficult to unambiguously identify in experiment and equally challenging to treat within standard theoretical models. Techniques such as resonant inelastic X-ray scattering (RIXS) depend critically on an accurate treatment of these hole-driven intermediate states, highlighting the need for many-body methods capable of explicitly resolving core-excited dynamics.

Many-body Green’s function approaches are natural candidates for this purpose, as they directly connect theory to experimental spectral functions. In the condensed-phase community, the $GW$ approximation of many-body perturbation theory has become the workhorse for quasiparticle energies~\cite{Hedin1965, Hybertsen1985,Golze2020,Caruso2012,Govoni2015, bintrim2021full, van2015gw, van2006quasiparticle,vlcek2017stochastic}. However, it is well known that $GW$ alone provides a poor description of satellite structures~\cite{guzzo2011valence,caruso2015band}. Cumulant expansions of the Green’s function ($GW$+C) address this deficiency by explicitly including higher-order correlations or vertex corrections, leading to an improved description of multiple satellites and better agreement with experiment~\cite{aryasetiawan1996multiple,lischner2013physical,kas2014cumulant,caruso2016gw,loos2024cumulant,kocklauner2025gw}. Nonetheless, their systematic improvability \textcolor{black}{is not fully established}, and their performance can vary depending on the choice of reference state (DFT or HF) and the specific form of the underlying $GW$ approximation~\cite{loos2024cumulant}.

Complementary to $GW$ approaches, coupled-cluster (CC) theory provides a systematically improvable framework encompassing Green’s function formalism~\cite{nooijen1992coupled,nooijen1993coupled,bhaskaran2016coupled,shee2019coupled,zhu2019coupled,peng2016coupled,peng2018green}, equation-of-motion~\cite{stanton1993equation, NooijenBartlett1995_EA, NooijenBartlett1995_IP}, and real-time formulations~\cite{monkhorst1977calculation,takahashi1986time,sverdrup2023time}. Specifically, in the time-dependent coupled-cluster (TD-CC) framework~\cite{monkhorst1977calculation,takahashi1986time,sverdrup2023time,kvaal2012ab,pedersen2020interpretation,huber2011explicitly,sato2018communication,nascimento2016linear}, the electronic wave function evolves according to an exponential ansatz that retains the desirable properties of the stationary CC theory while extending its applicability to nonequilibrium and time-resolved phenomena. Previous developments have demonstrated the versatility of TD-CC in evaluating excitation energies~\cite{takahashi1986time} and core-excitation spectra~\cite{skeidsvoll2020time,park2019equation} and computing response properties~\cite{dalgaard1983some,coriani2016molecular}. Further extensions have incorporated relativistic effects~\cite{koulias2019relativistic}, finite-temperature~\cite{white2018time}, reduced-scaling implementations~\cite{peyton2023reduced}, and adaptive numerical integration techniques~\cite{vila2025efficient,wang2022accelerating}, \textcolor{black}{as well as} extension to multi-reference formulations~\cite{mosquera2025time}. In addition to TD-CC formulations, other real-time \textit{ab initio} approaches have also been developed to simulate electronic dynamics~\cite{lopata2011modeling,li2020real,sato2013time,peng2018simulating,klamroth2003laser,peyton2023tailoring}.

Building on the TD-CC Green’s function work of Sch\"onhammer and Gunnarsson~\cite{schonhammer1978time}, the real-time equation-of-motion coupled-cluster (RT-EOM-CC) cumulant Green’s function approach extends this framework within a coupled-cluster formalism. Here, the Green's function formalism employs an exponential cumulant representation in which a core state is created and its subsequent time evolution is followed to obtain Green’s functions for core-level spectra~\cite{rehr2020equation, vila2022real, vila2024rt, vila2025efficient}. In particular, a TD-CC ansatz is employed for propagating the correlated ionized $(N-1)$-electron state, which naturally leads to a nonperturbative cumulant representation of the one-particle Green’s function, which has the similar conceptual ground as cumulant-based Green’s function approaches, but with the advantage of \textcolor{black}{systematically improvable} CC correlation treatments. RT-EOM-CCSD cumulant Green's function approach has been applied to both core and valence ionization, where it reproduces quasiparticle energies with sub-eV accuracy and captures many-body satellites in good agreement with experiment and high-level benchmarks~\cite{vila2024rt,vila2025efficient}.  

Despite these advances, the original TD-CC ansatz for the correlated ionized state was limited by its reliance on a single exponential parametrization in the $(N-1)$-electron space. In this formulation, the core-hole state was generated on top of a mean-field reference, \textcolor{black}{which neglects} correlations carried over from the correlated $N$-electron ground state. To overcome this limitation, a time-dependent double coupled-cluster (TD-dCC) ansatz has recently been proposed~\cite{peng2024exploring}. In the TD-dCC formulation, the correlated ionized state is parametrized as a product of two CC exponentials, one built from the correlated $N$-electron ground state and the other from the $(N-1)$-electron sector. This dual-exponential form accounts for secondary, hole-mediated excitations induced by core-hole creation, allowing the ionized state to couple with ground-state correlations. Similar dCC–type approaches have been explored previously in coupled-cluster frameworks, mainly within time-independent formulations~\cite{meissner1998fock,nooijen1996many,nooijen1997similarity, bauman2019downfolding,tribedi2020formulation, bauman2022coupled}.

In this work, we investigate a more cost-effective approximation hierarchy for the approach introduced in Ref.~\citenum{peng2024exploring} to be applied to molecular systems, and also investigate the performance of quantum algorithms for computing Green’s functions. We demonstrate that the TD-dCC ansatz and its new approximations in the RT-EOM-CC cumulant Green's function approach provide accurate quasiparticle ionization energies while significantly improving the description of satellite features in molecular spectral functions in comparison with all the previous single-exponential approaches. We also perform a component analysis of the spectral function, which allows us to distinguish hole-mediated transitions in the TD-dCC hierarchy. These transitions, arising from the coupling between ground-state and ionized-state correlation effects through the core-hole, are absent in single-exponential CC approaches and highlight the capability of the TD-dCC approach and its approximations to capture correlated excitation pathways that contribute to both quasiparticle peaks and satellites.

While the TD-dCC ansatz can capture many-body effects, it can also miss satellite transitions associated with multiple electrons that go beyond the excitation manifold used. On the other hand, the inclusion of higher-order excitations can become prohibitively expensive due to steep scaling and memory demands (e.g., CC with triples excitation scales as $N^8$). In this regard, quantum algorithms offer a promising alternative. Through directly encoding the many-body Hamiltonian, the quantum algorithms can potentially alleviate classical scaling limitations and provide access to correlated states with more favorable resource scaling~\cite{alexeev2025perspectivequantumcomputingapplications}. In particular, the \textcolor{black}{use of} fault-tolerant quantum simulations potentially offer utilities that go beyond the reach of conventional computing~\cite{alexeev2025perspectivequantumcomputingapplications}. Hence we also analyze a fault-tolerant quantum algorithm for obtaining the time-dependent $N-1$ correlation for computing the Green's function of molecular systems using block encoding and quantum signal processing (QSP)/quantum singular value transformation (QSVT) techniques.

The paper is organized as follows. Section~\ref{sec:theory} outlines the theoretical framework, beginning with a brief review of the formulation of the one-particle Green’s function for a deep core-hole, the RT-EOM-CC cumulant Green's function approach, and the TD-CC ansatz, followed by the development of the TD-dCC and approximate TD-dCC ans\"{a}atzes. The subsequent sections describe the computation of quasiparticle weights and the component analysis of the TD-dCC spectral function. Section~\ref{sec:compu_detail} summarizes the computational details. Section~\ref{sec:results} presents numerical results for the single-impurity Anderson model (SIAM), H$_2$O at both equilibrium and stretched geometries, and CH$_4$ molecular systems, and discusses the implementation of the quantum algorithm based on QSP/QSVT. Finally, Section~\ref{sec:conclusion} provides concluding remarks and future perspectives.

%%%%%%%%%%%%%%%%%%%%%%%%%%%%

\section{Theory}\label{sec:theory}

%-------------------------

\subsection{One-particle Green’s function}

The retarded one-particle Green’s function for a deeply occupied orbital $c$ is given by 
\begin{align}
G^R_c(t) &= -i\Theta(t)\langle \Psi^{(N)} | [a_c(t), a_c^\dagger(0)]_+ | \Psi^{(N)} \rangle \notag \\
&\approx -i\Theta(t)e^{-iE^{(N)}_g t}\langle \Psi^{(N-1)}_c | e^{i\hat{H}t} | \Psi^{(N-1)}_c \rangle.
\label{eq:GF}
\end{align}
{\color{black}
Here, we assume \( a_c^\dagger \lvert \Psi^{(N)} \rangle \approx 0 \), as the spin-orbital \( c \) is already occupied in the \( N \)-electron ground state. 
The function \( \Theta(t) \) is the Heaviside step function.  
The state \( \lvert \Psi^{(N-1)}_c \rangle = a_c \lvert \Psi^{(N)} \rangle \) corresponds to the instantaneous \((N-1)\)-electron configuration obtained by removing an electron from spin-orbital  $c$  in the \( N \)-electron ground state \( \lvert \Psi^{(N)} \rangle \).  
The time-dependent annihilation and creation operators in the Heisenberg picture are defined as  
\[
a_c(t) = e^{iHt} a_c e^{-iHt}, \qquad
a_c^\dagger(t) = e^{iHt} a_c^\dagger e^{-iHt}.
\]
}

Within the coupled-cluster (CC) framework, the ground-state wavefunction is represented by the exponential ansatz
\begin{align}
|\Psi^{(N)}\rangle &= e^{T^{(N)}} |\phi^{(N)}_0\rangle, \label{eq:ccket} \\
\langle \Psi^{(N)}| &= \langle \phi^{(N)}_0| \left( 1 + \Lambda^{(N)} \right) e^{-T^{(N)}}, \label{eq:ccbra}
\end{align}
where \(T^{(N)}\) is the cluster operator generating particle–hole excitations from the reference determinant \(|\phi^{(N)}_0\rangle\), and \(\Lambda^{(N)}\) is the corresponding de-excitation operator. Due to the non-Hermitian nature of the CC parametrization, the left and right ground states are biorthogonal. The coupled-cluster Green’s function (CCGF) formalism extends the CC framework to compute spectral and response properties by expressing the one-particle Green’s function in terms of CC excitation/de-excitation operators~\cite{nooijen1992coupled,nooijen1993coupled,stanton1993equation,bhaskaran2016coupled,peng2018green,shee2019coupled,backhouse2022constructing}.

\textcolor{black}{For clarity, we note here that the index $c$ consistently denotes the specific core spin-orbital from which an electron is removed to generate the \((N{-}1)\)-electron state. This convention is used throughout the TD-CC and TD-dCC formulations below.}

%---------------------------------

\subsection{RT-EOM-CC cumulant Green's function approach and TD-CC ansatz}

The RT-EOM-CC cumulant Green's function approach provides an alternative approach where a core-hole state is created and its subsequent time evolution is followed to obtain Green’s functions for core-level spectra~\cite{rehr2020equation, vila2022real, vila2024rt, vila2025efficient}. The detailed mathematical framework of the approach has been presented in earlier work~\cite{vila2022real,rehr2020equation,vila2024rt}. Here, we summarize only the essential aspects relevant to the present study. The framework \textcolor{black}{adopts a complementary strategy to the traditional CCGF formulation} by introducing a time-dependent coupled-cluster parametrization of an approximate correlated $(N-1)$-electron state, whose propagation yields the Green’s function in cumulant form~\cite{vila2022real}.  

The TD-CC ansatz for the correlated \((N-1)\)-electron core-hole state is
\begin{equation}
|\Psi^{(N-1)}_c(t)\rangle = N_c(t)\, e^{T^{(N-1)}(t)}|\phi^{(N-1)}_0\rangle,
\label{eq:cc-ansatz}
\end{equation}
where \(N_c(t)\) is a normalization factor and \(T^{(N-1)}(t)\) is the cluster operator in the \((N-1)\)-electron space. $|\phi^{(N-1)}_0\rangle$ denotes the \((N-1)\)-electron reference state. The corresponding equations of motion are
\begin{align}
\small
-\,i \frac{\partial}{\partial t}\ln N_c(t) &= \langle \phi^{(N-1)}_0|\bar{H}(t)|\phi^{(N-1)}_0\rangle \notag \\
&=  E^{(N-1)}_{CC}(t), \label{eq:EOM-Nc} \\
-\,i \frac{\partial}{\partial t}\, t^{(N-1)}_n(t) &= \langle n^{(N-1)}|\bar{H}(t)|\phi^{(N-1)}_0\rangle, \label{eq:EOM-t}
\end{align}
with the similarity-transformed Hamiltonian \(\bar{H}(t)=e^{-T^{(N-1)}(t)}He^{T^{(N-1)}(t)}\).  

The corresponding retarded Green’s function takes the form
\begin{align}
G^R_c(t) &= -i\Theta(t)e^{-iE^{(N)}_{CC}t} N_c(t) \notag \\
&= -i\Theta(t)e^{-i\Delta E_{CC}(t)t},
\label{eq:GF-cc}
\end{align}
where
\begin{equation}
\Delta E_{CC}(t) = E^{(N)}_{CC} - [E^{(N-1)}_{CC}]_t
\label{eq:deltaE}
\end{equation}
and \([E^{(N-1)}_{CC}]_t\) denotes the averaged \((N-1)\)-electron CC energy over the time $t$  
\begin{align}
[E^{(N-1)}_{\text{CC}}]_t = \frac{1}{t}\int_0^t E^{(N-1)}_{\text{CC}}(\tau)\,d\tau .
\end{align}

\textcolor{black}{In contrast to the static CCGF approach, the RT-EOM-CC cumulant Green's function formalism captures dynamical orbital relaxation and shake-up processes through the explicit time dependence of the \((N{-}1)\)-electron cluster amplitudes. However, as discussed below, its single-exponential parametrization limits the extent to which ground-state correlations from the $N$-electron sector influence the ionized dynamics.}

%---------------------------------

\begin{figure*}
\centering
\includegraphics[width=\linewidth]{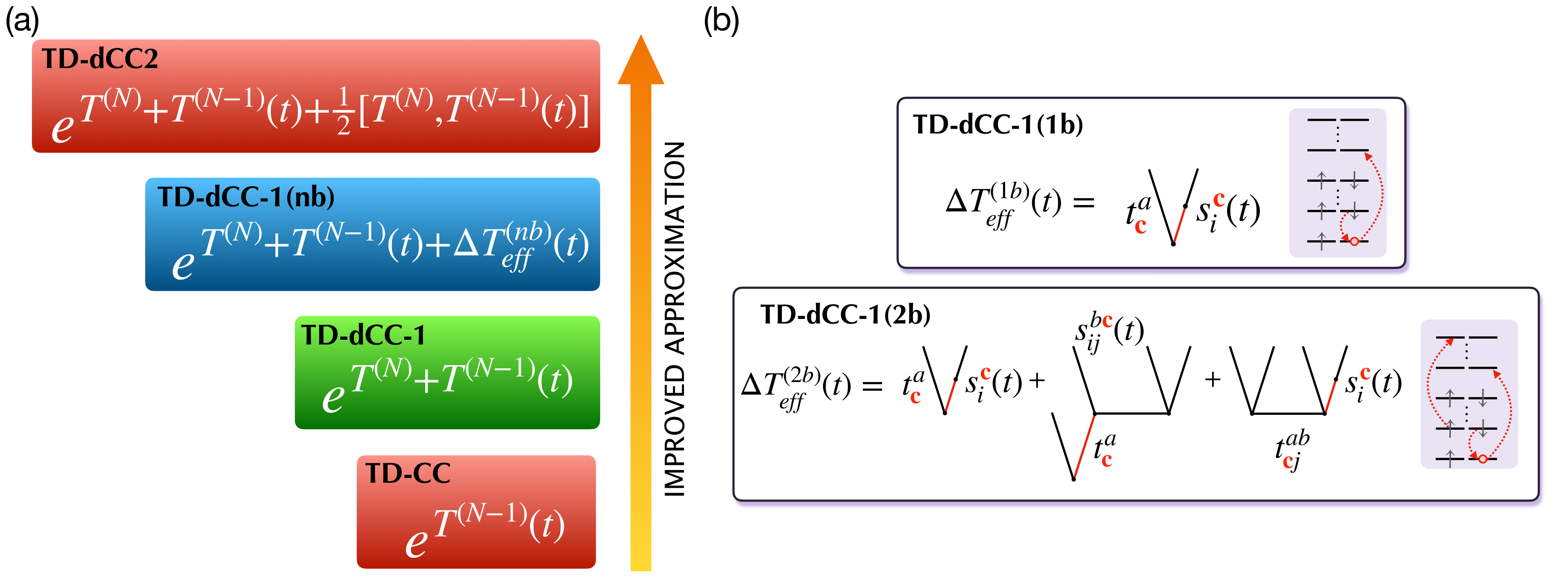}
\caption{(a) Systematic hierarchy of approximations to the time-dependent many-body ans\"{a}tzes employed in the RT-EOM-CC cumulant Green's function approach. (b) Diagrams included in the commutator expansion corresponding to the effective correction $\Delta T^{(nb)}_{\rm eff}$ in the approximate TD-dCC-1($n$b) ansatz. The $s_{i}^{a}(t)$ and $s_{ij}^{ab}(t)$ correspond to the time-dependent single and double cluster amplitudes in the $N{-}1$ sector, while $t_{i}^{a}$ and $t_{ij}^{ab}$ correspond to the $N$-electron cluster amplitudes. The index $c$ denotes the ionized core spin-orbital. \textcolor{black}{This hierarchy illustrates how progressively including commutator terms restores $N$–$(N{-}1)$ correlation pathways important for satellite formation.}}
\label{fig:dcc1nb}
\end{figure*}

\subsection{TD-dCC ansatz}\label{sec:dcc_nb}

While the TD-CC ansatz formally enables the propagation of $(N-1)$-electron dynamics, its description of the core-hole state is restricted to the correlation based on a mean-field reference $|\phi^{(N-1)}_0\rangle$. Therefore, it overlooks the feedback of correlations from the correlated $N$-electron ground state. As a result, the approach cannot guarantee convergence to the exact one-particle Green’s function, in contrast to the CCGF formalism, where correlations between the $N$- and $(N\!-\!1)$-electron sectors are approached consistently~\cite{peng2018green,peng2024exploring}. To address this issue, the TD-dCC ansatz was recently introduced~\cite{peng2024exploring}, which simultaneously incorporates correlations from both the $N$- and $(N-1)$-electron sectors. 

The TD-dCC wavefunction is given by  
\begin{equation}
    |\Psi^{(N-1)}_c(t)\rangle = \tilde{N}_c(t)\, e^{T^{(N)}} e^{T^{(N-1)}(t)}|\phi^{(N-1)}_0\rangle,
\label{eq:dCC}
\end{equation}
where $T^{(N)}$ accounts for the correlation effects in the $N$-electron ground state, and $T^{(N-1)}(t)$ describes the real-time dynamics within the ionized $(N-1)$-electron manifold. This ansatz extends the original TD-CC ansatz \eqref{eq:cc-ansatz} by incorporating hole-mediated higher-order excitations, thereby providing a systematically improvable path toward the exact Green’s function. 

\textcolor{black}{Conceptually, the dual-exponential structure allows the ionized-state dynamics to couple directly to pre-existing $N$-electron correlations---an effect that is inaccessible in single-exponential TD-CC ansatz and is essential for capturing many-body satellite formation and shake-up pathways.}
The corresponding equations of motion are
\begin{align}
-\,i \frac{\partial}{\partial t}\ln \tilde{N}_c(t) &= \langle \phi^{(N-1)}_0|\bar{\bar{H}}(t)|\phi^{(N-1)}_0\rangle \notag \\
&=  E^{(N-1)}_{dCC}(t), \label{eq:dCC-Nc} \\
-\,i \frac{\partial}{\partial t}\, t^{(N-1)}_n(t) &= \langle n^{(N-1)}|\bar{\bar{H}}(t)|\phi^{(N-1)}_0\rangle, \label{eq:dCC-t}
\end{align}
with the double similarity-transformed Hamiltonian
\begin{equation}
\bar{\bar{H}}(t) = e^{-T^{(N-1)}(t)}e^{-T^{(N)}} H\, e^{T^{(N)}}e^{T^{(N-1)}(t)}.
\label{eq:dblH}
\end{equation}
The Green’s function in this new framework becomes
\begin{equation}
G^R_c(t) = -i\Theta(t)e^{-iE^{(N)}_{CC}t}\,\tilde{N}_c(t)\tilde{O}(t),
\label{eq:GF-dCC}
\end{equation}
where  \(\tilde{O}(t)\) is the time-dependent overlap function. In the previous RT-EOM-CC cumulant Green's function formalism (see Eq.~\eqref{eq:GF-cc}), this overlap is approximated to one. For the TD-dCC ansatz, we compute this as 
\begin{equation}
    \small
    \tilde{O}(t) = \langle \phi_0^{(N)} | (1 + \Lambda^{(N)}) e^{-T^{(N)}} a_c^\dagger 
| \Psi_c^{(N-1)}(t) \rangle .
\label{eq:ovlp}
\end{equation}
This TD-dCC construction enables the real-time coupled-cluster Green's function to reproduce the exact Green’s function in the expansion limit and improves the description of satellite structures in correlated spectra as shown in Ref.~\citenum{peng2024exploring}. 

\textcolor{black}{We emphasize that the overlap factor $\tilde{O}(t)$ carries the explicit coupling between $N$- and $(N{-}1)$-electron sectors and is responsible for the emergence of hole-mediated excitations analyzed later in Sec.~\ref{sec:comp}.}

%---------------------------------

\subsubsection{Approximate TD-dCC ans\"{a}tzes}

In the TD-dCC framework, the use of two exponential operators results in a double similarity transformation when deriving the EOMs. This construction introduces additional non-linear terms that may hinder stable numerical time propagation. \textcolor{black}{To retain the physical content of TD-dCC while simplifying its algebraic structure, we derive practical approximations based on truncating the BCH expansion of $e^{T^{(N)}}e^{T^{(N-1)}(t)}$.} 
Two such approximations were introduced in Ref.~\cite{peng2024exploring}:

\begin{itemize}
\item In the \textbf{TD-dCC-1} approximation, we employ a first-order truncation,
\begin{equation}
|\Psi^{(N-1)}_c(t)\rangle \approx \tilde{N}_c(t)\, 
e^{T^{(N)} + T^{(N-1)}(t)} |\phi^{(N-1)}_0\rangle.
\end{equation}
\textcolor{black}{This retains the essential $N$–$(N{-}1)$ coupling while avoiding higher commutators.}

\item In the \textbf{TD-dCC-2} approximation, we retain the first commutator term,
\begin{align}
\small
&|\Psi^{(N-1)}_c(t)\rangle \approx \tilde{N}_c(t)\, \times \notag \\
&~~~~e^{T^{(N)} + T^{(N-1)}(t)  + \tfrac{1}{2}[T^{(N)}, T^{(N-1)}(t)]} 
|\phi^{(N-1)}_0\rangle. \label{eq:dcc2}
\end{align}
\textcolor{black}{The commutator introduces higher-body terms and improves accuracy but increases computational cost.}
\end{itemize}

Both TD-dCC-1 and TD-dCC-2 maintain the single-exponential structure of the coupled-cluster ansatz while incorporating a limited number of static $N$-electron $T$-amplitudes. Solving TD-dCC-2 requires explicit evaluation of the commutator between the $N$- and $(N-1)$-electron operators, which \textcolor{black}{introduces additional cost} compared to the TD-dCC-1 approximation.

%---------------------------------

\subsubsection{Systematic n-body corrections to the TD-dCC-1 ansatz}

The progression from the original TD-CC ansatz~\eqref{eq:cc-ansatz} to TD-dCC-2~\eqref{eq:dcc2} can be viewed as a systematic perturbative hierarchy, where the original CC formalism serves as the zeroth-order approximation, while TD-dCC-1 and TD-dCC-2 represent higher-order corrections in the BCH commutator expansion.

To provide finer control over this expansion, we introduce an intermediate correction scheme that selectively retains commutator terms based on their excitation rank. This leads to the \emph{time-dependent double coupled cluster $n$-body approximation}, denoted as TD-dCC-1($n$b), which systematically includes commutator corrections up to the $n$-body level while discarding higher-order terms that would generate excitations beyond the target manifold.  
\textcolor{black}{This approach interpolates smoothly between TD-dCC-1 (cheapest) and TD-dCC-2 (most accurate), while keeping the cost comparable to TD-CC for the singles–doubles (SD) truncation.}

The ansatz for the TD-dCC-1($n$b) is
\begin{equation}
\small
|\Psi^{(N-1)}_c(t)\rangle \approx \tilde{N}_c(t)\ 
e^{T^{(N)} + T^{(N-1)}(t)  + \Delta T^{(nb)}_{\text{eff}}(t)} 
|\phi^{(N-1)}_0\rangle. \label{eq:dcc2_new}
\end{equation}
For a SD excitation manifold, the highest admissible excitation rank in the commutator expansion is \( n_{\max} = 2 \). Therefore, triple and higher excitations generated from commutators involving \(T^{(N)}\) and \(T^{(N-1)}\) are discarded. Similar reasoning applies to singles-doubles-triples or higher manifolds.

\paragraph{TD-dCC-1(1b): Single-body corrections.} 
In the one-body variant, we retain only the singles sectors in the commutator, yielding
\begin{align}
\Delta T^{(1b)}_{\text{eff}}(t) 
&= \frac{1}{2} [T^{(N)}_1, T^{(N-1)}_1(t)] \nonumber \\
&= \sum_{i \neq \mathbf{c}} \sum_a \left( \tfrac{1}{2} t_\mathbf{c}^a s_i^{\mathbf{c}}(t)\right) a_a^\dagger  a_i  .
\end{align}
\textcolor{black}{This term captures hole-mediated singles-level couplings between $N$- and $(N{-}1)$-electron sectors.}

\paragraph{TD-dCC-1(2b): Two-body corrections.} 
The two-body approximation incorporates mixed commutators involving singles and doubles:
\begin{align}
\Delta T^{(2b)}_{\text{eff}} 
  &= \tfrac{1}{2} \Big(
    [T^{(N)}_1, T^{(N-1)}_1(t)]
    + [T^{(N)}_2, T^{(N-1)}_1(t)]  \notag \\
  &~~~~~~ + [T^{(N)}_1, T^{(N-1)}_2(t)]
  \Big).
\end{align}
\textcolor{black}{These terms generate effective doubles-level couplings and enhance the description of multi-electron shake-up processes.}
Higher commutators (e.g., $[T^{(N)}_2, T^{(N-1)}_2]$) are omitted because they generate excitations beyond the SD space.

\textcolor{black}{Overall, the TD-dCC-1($n$b) hierarchy retains the computational scaling of TD-CC while incorporating the essential $N$–$(N{-}1)$ correlation pathways needed for accurate satellite descriptions, as illustrated earlier in Figure~\ref{fig:dcc1nb}.}

%---------------------------------

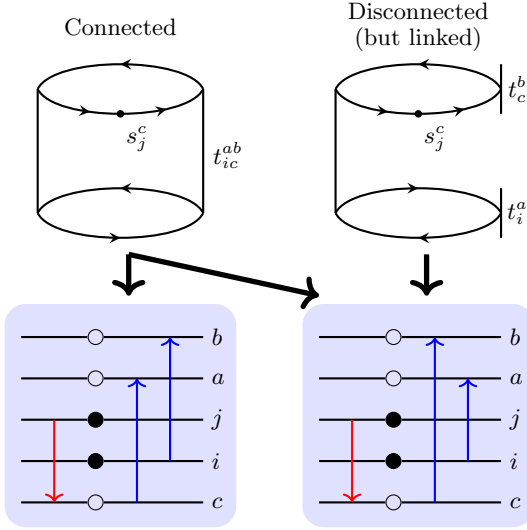
\begin{figure}
\centering
\begin{tikzpicture}[scale=1.1, every node/.style={font=\small}]

\def\tlx{0.4}   % top left x coord
\def\tly{1.5}   % top left y coord
\def\trx{4}     % top right x coord
\def\try{1.5}   % top right y coord
\def\blx{0.2}   % bottom left x coord
\def\bly{-1.5}    % bottom left y coord
\def\brx{3.8}   % bottom right x coord
\def\bry{-1.5}    % bottom right y coord

\draw[line width=2pt, ->] (\tlx+1.1,\tly-2) -- (\blx+1.3,\bly+0.5);

\draw[line width=2pt, ->] (\trx+1.1,\try-2) -- (\brx+1.3,\bry+0.5);

\draw[line width=2pt, ->] (\tlx+1.1,\tly-2) -- (\brx,\bry+0.5);

\draw[rounded corners=8pt, draw=none,
      fill=blue!40, 
      fill opacity=0.3] 
    (\blx-0.2,\bly+0.4) rectangle (\blx+2.6,\bly-2.3);

\draw[rounded corners=8pt, draw=none,
      fill=blue!40, 
      fill opacity=0.3] 
    (\brx-0.2,\bry+0.4) rectangle (\brx+2.6,\bry-2.3);

\node at (\tlx+1,\tly+0.75) {Connected};
\node at (\trx+1,\try+0.95) {Disconnected};
\node at (\trx+1,\try+0.6) {(but linked)};

% Curved top and bottom
\draw[
  thick,
  postaction={
    decorate,
    decoration={
      markings,
      mark=at position 0.5 with {\arrow{stealth}}
    }
  }
] (\tlx+2,\tly) .. controls (\tlx+1.8,\tly+0.4) and (\tlx+0.2,\tly+0.4) .. (\tlx,\tly);
\draw[
  thick,
  postaction={
    decorate,
    decoration={
      markings,
      mark=at position 0.33 with {\arrow{stealth}},
      mark=at position 0.76 with {\arrow{stealth}}
    }
  }
] (\tlx,\tly) .. controls (\tlx+0.2,\tly-0.4) and (\tlx+1.8,\tly-0.4) .. (\tlx+2,\tly);
\draw[
  thick,
  postaction={
    decorate,
    decoration={
      markings,
      mark=at position 0.5 with {\arrow{stealth}}
    }
  }
] (\tlx+2,\tly-1.5) .. controls (\tlx+1.8,\tly-1.1) and (\tlx+0.2,\tly-1.1) .. (\tlx,\tly-1.5);
\draw[
  thick,
  postaction={
    decorate,
    decoration={
      markings,
      mark=at position 0.5 with {\arrow{stealth}}
    }
  }
] (\tlx,\tly-1.5) .. controls (\tlx+0.2,\tly-1.9) and (\tlx+1.8,\tly-1.9) .. (\tlx+2,\tly-1.5);

% Vertical sides
\draw[thick] (\tlx,\tly) -- (\tlx,\tly-1.5);
\draw[thick] (\tlx+2,\tly) -- (\tlx+2,\tly-1.5);

% Internal dot
\node[circle,fill,inner sep=1pt] at (\tlx+1,\tly-0.3) {};

% Labels
\node at (\tlx+1.2,\tly-0.6) {$s_j^c$};
\node at (\tlx+2.3,\tly-0.8) {$t_{ic}^{ab}$};

%%%%%%%%%%%%%%%%%%%

% Curved top and bottom
\draw[
  thick,
  postaction={
    decorate,
    decoration={
      markings,
      mark=at position 0.5 with {\arrow{stealth}}
    }
  }
] (\trx+2,\try) .. controls (\trx+1.8,\try+0.4) and (\trx+0.2,\try+0.4) .. (\trx,\try);
\draw[
  thick,
  postaction={
    decorate,
    decoration={
      markings,
      mark=at position 0.33 with {\arrow{stealth}},
      mark=at position 0.76 with {\arrow{stealth}}
    }
  }
] (\trx,\try) .. controls (\trx+0.2,\try-0.4) and (\trx+1.8,\try-0.4) .. (\trx+2,\try);
\draw[
  thick,
  postaction={
    decorate,
    decoration={
      markings,
      mark=at position 0.5 with {\arrow{stealth}}
    }
  }
] (\trx,\try-1.5) .. controls (\trx+0.2,\try-1.1) and (\trx+1.8,\try-1.1) .. (\trx+2,\try-1.5);
\draw[
  thick,
  postaction={
    decorate,
    decoration={
      markings,
      mark=at position 0.5 with {\arrow{stealth}}
    }
  }
] (\trx+2,\try-1.5) .. controls (\trx+1.8,\try-1.9) and (\trx+0.2,\try-1.9) .. (\trx,\try-1.5);

% Vertical sides
\draw[thick] (\trx,\try) -- (\trx,\try-1.5);
\draw[thick] (\trx+2,\try+0.3) -- (\trx+2,\try-0.3);
\draw[thick] (\trx+2,\try-1.2) -- (\trx+2,\try-1.8);

% Internal dot
\node[circle,fill,inner sep=1pt] at (\trx+1,\try-0.3) {};

% Labels
\node at (\trx+1.2,\try-0.6) {$s_j^c$};
\node at (\trx+2.2,\try) {$t_{c}^{b}$};
\node at (\trx+2.2,\try-1.5) {$t_{i}^{a}$};

%%%%%%%%%%%%%%%%%%%%%%%%%%%%%%%%%%

% Energy levels
\foreach \y/\label in {\bly/$b$,\bly-0.5/$a$,\bly-1/$j$,\bly-1.5/$i$,\bly-2/$c$} {
    \draw[thick] (\blx,\y) -- (\blx+0.8,\y);
    \draw[thick] (\blx+1,\y) -- (\blx+2.2,\y);
    \node[right] at (\blx+2.2,\y) {\label};
}

% Excitation arrows (blue)
\draw[blue, thick, ->] (\blx+1.4,\bly-2) -- (\blx+1.4,\bly-0.5);
\draw[blue, thick, ->] (\blx+1.8,\bly-1.5) -- (\blx+1.8,\bly);

% De-excitation arrow (red)
\draw[red, thick, ->] (\blx+0.4,\bly-1) -- (\blx+0.4,\bly-2);

\node[draw,circle,inner sep=2pt] at (\blx+0.9,\bly-2) {};
\node[draw,circle,fill,inner sep=2pt] at (\blx+0.9,\bly-1.5) {};
\node[draw,circle,fill,inner sep=2pt] at (\blx+0.9,\bly-1) {};
\node[draw,circle,inner sep=2pt] at (\blx+0.9,\bly-0.5) {};
\node[draw,circle,inner sep=2pt] at (\blx+0.9,\bly) {};

%%%%%%%%%%%%%%%%%%%%%%%%%%%%%%%%%%

% Energy levels
\foreach \y/\label in {\bry/$b$,\bry-0.5/$a$,\bry-1/$j$,\bry-1.5/$i$,\bry-2/$c$} {
    \draw[thick] (\brx,\y) -- (\brx+0.8,\y);
    \draw[thick] (\brx+1,\y) -- (\brx+2.2,\y);
    \node[right] at (\brx+2.2,\y) {\label};
}

% Excitation arrows (blue)
\draw[blue, thick, ->] (\brx+1.4,\bry-2) -- (\brx+1.4,\bry);
\draw[blue, thick, ->] (\brx+1.8,\bry-1.5) -- (\brx+1.8,\bry-0.5);

% De-excitation arrow (red)
\draw[red, thick, ->] (\brx+0.4,\bry-1) -- (\brx+0.4,\bry-2);

\node[draw,circle,inner sep=2pt] at (\brx+0.9,\bry-2) {};
\node[draw,circle,fill,inner sep=2pt] at (\brx+0.9,\bry-1.5) {};
\node[draw,circle,fill,inner sep=2pt] at (\brx+0.9,\bry-1) {};
\node[draw,circle,inner sep=2pt] at (\brx+0.9,\bry-0.5) {};
\node[draw,circle,inner sep=2pt] at (\brx+0.9,\bry) {};

\end{tikzpicture}
\caption{Connected and disconnected (but linked) diagrams contributing to the overlap $\tilde{O}^{\text{HM}}_{(i,j,a,b)}(t)$, where the HM transitions are described through the TD-dCC ans\"{a}tzes. The core hole is located in spin-orbital $c$. The connected diagram corresponds to two symmetry-equivalent HM transitions due to the antisymmetry of $t_{ic}^{ab}$ with respect to virtual indices $a$ and $b$, whereas the disconnected diagram (linked to the left eigenvector) corresponds to a specific HM transition. \textcolor{black}{These diagrams reveal the algebraic origin of HM coupling channels captured by the TD-dCC framework.}}
\label{fig:HM transfer}
\end{figure}

\subsection{Component Analysis of TD-dCC Spectral Function}\label{sec:comp}

To characterize individual excitation contributions to the spectral function and identify hole-mediated (HM) transitions, we perform an elaborated component analysis by decomposing the time-dependent overlap function $\tilde{O}(t)$ according to the excitation structure of the TD-dCC ans\"{a}tzes. A key advantage of the TD-dCC framework is its ability to capture hole-mediated transitions, excitation pathways that couple $N$-electron ground-state correlations to $(N-1)$-electron ionized-state dynamics through the core-hole. These transitions arise from cross-terms between $T^{(N)}$ and $T^{(N-1)}(t)$ operators and are absent in the original single-exponential CC ansatz. 

\textcolor{black}{This decomposition provides a physically transparent way to determine whether specific spectral features originate from direct $(N{-}1)$-electron excitations or from correlation-driven, hole-mediated processes activated by the core-hole.}

For the single-exponential approximations TD-dCC-1 and TD-dCC-1($n$b), the overlap function is
\begin{equation}
    \tilde{O}(t) = \langle \Psi^{(N)} | a_c^\dagger \, e^{T + S + \Delta} \, a_c | \phi_0^{(N)} \rangle,
    \label{eq:overlap_master}
\end{equation}
where we denote $T \equiv T^{(N)}$ as the time-independent $N$-electron cluster operator, $S \equiv T^{(N-1)}(t)$ the real-time cluster operator for the ionized state, and $\Delta \equiv \Delta T_{\text{eff}}(t)$ the $n$-body commutator corrections.
We perform a balanced truncation of the Taylor expansion of the exponential operator in Eq.~\eqref{eq:overlap_master} to the second order:
\begin{align}
e^{T + S + \Delta} \approx &\, 1 + T + S + \tfrac{1}{2}(T^2 + S^2 + TS + ST) \notag \\
&+ \Delta + \tfrac{1}{2}\{T + S, \Delta\} + \mathcal{O}(\Delta^2).
\label{eq:balanced_truncation}
\end{align}
\textcolor{black}{This controlled truncation allows us to systematically group terms according to their physical origin while retaining all contributions relevant within the SD excitation manifold.}

We then classify contributions to the time-dependent overlap $\tilde{O}(t)$ into two physically distinct categories,
\begin{equation}
\tilde{O}(t) 
= \sum_{\mu} \tilde{O}^{\text{direct}}_{\mu}(t) 
+ \sum_{\nu} \tilde{O}^{\text{HM}}_{\nu}(t)~.
\label{eq:O_decomp}
\end{equation}
Here, the first part of Eq.~\eqref{eq:O_decomp} collects the direct transitions in the ionized $(N\!-\!1)$-electron excitation manifold, 
\begin{equation}
\tilde{O}^{\text{direct}}_{\mu}(t)
= \langle \Psi^{(N)} | 
a_c^\dagger \, \kappa^{(N-1)}_{\mu}(t)\, \mathcal{E}_{\mu}\, a_c 
| \phi_0^{(N)} \rangle ,
\label{eq:O_direct}
\end{equation}
where $\mu$ indexes single excitations $\mu=(i,a)$ and double excitations $\mu=(i,j,a,b)$, and $\mathcal{E}_{\mu}$ denotes the corresponding excitation operator ($a_a^\dagger a_i$ or $a_a^\dagger a_b^\dagger a_j a_i$). 
The second part of Eq.~\eqref{eq:O_decomp} contains the HM transitions resolved by the TD-dCC ans\"{a}tzes, i.e., the cross terms involving $T$ and $S$, such as $TS$ and commutator corrections from $\Delta$ involving the core-hole index $c$, that do not appear in the original single-exponential CC formulation:
\begin{equation} 
\tilde{O}^{\text{HM}}_{\nu}(t)
= \langle \Psi^{(N)} | 
a_c^\dagger \, \tilde{\Omega}_{\nu}(t)\, \mathcal{E}_{\nu}\, a_c 
| \phi_0^{(N)} \rangle ,
\label{eq:O_HM}
\end{equation}
where $\tilde{\Omega}_{\nu}(t)$ are effective amplitudes built from nonlinear products of $T$, $S$, and $\Delta$, and $\nu$ labels the excitation channels generated by these couplings. 
\textcolor{black}{These HM amplitudes encode how the core-hole activates correlated transitions that are inaccessible to the $(N{-}1)$-sector alone---precisely the processes responsible for satellite formation and multi-electron shake-up.}

The structure of $\tilde{\Omega}_{\nu}(t)$ varies across approximation levels:
\begin{itemize}
\item \textbf{TD-dCC-1}: Only the $\tfrac{1}{2}TS(t)$ cross-term contributes:
\begin{align}
    \tilde{\Omega}_{i}^{a}(t) &= \tfrac{1}{2} t_{c}^{a}\, s_i^{c}(t) , \label{eq:dcc-1-comp}\\
    \tilde{\Omega}_{ij}^{ab}(t) &= \tfrac 1 8 \bigg(
    t_c^a s_{ij}^{cb}(t) + t_c^b s_{ij}^{ac}(t) + t_{cj}^{ab} s_i^c(t) \notag \\
    &~~~~~~~~ + t_{ic}^{ab} s_j^c(t) \bigg) ;
\end{align}
\item \textbf{TD-dCC-1($1$b)}: Additional contributions from $\Delta_1$ and $\tfrac{1}{2}\{T + S, \Delta_1\}$ appear (with $\Delta_1$ the singles commutator):
\begin{align}
    \tilde{\Omega}_{i}^{a}(t) &=  t_{c}^{a}\, s_i^{c}(t) , \\
    \tilde{\Omega}_{ij}^{ab}(t) &= \tfrac 1 8 \bigg(
    t_c^a s_{ij}^{cb}(t) + t_c^b s_{ij}^{ac}(t) + 
    t_{cj}^{ab} s_i^c(t) + t_{ic}^{ab} s_j^c(t) \nonumber \\
    & \qquad \qquad  + 4~t_i^a t_c^b s_j^c(t) 
      + 4~s_i^a(t)t_c^b s_j^c(t) \bigg) ;
\end{align}
\item \textbf{TD-dCC-1(2b)} and \textbf{TD-dCC-2}: All contributions to the single commutator $\Delta$ remain within the SD excitation manifold:
\begin{align}
    \tilde{\Omega}_{i}^{a}(t) &=  t_{c}^{a}\, s_i^{c}(t) , \\
    \tilde{\Omega}_{ij}^{ab}(t) &= 
    \tfrac 1 4 \bigg(t_c^a s_{ij}^{cb}(t) + t_c^b s_{ij}^{ac}(t) + 
    t_{cj}^{ab} s_i^c(t) + t_{ic}^{ab} s_j^c(t) \nonumber \\
    & \qquad \qquad + 2~t_i^a t_c^b s_j^c(t) 
      + 2~s_i^a(t)t_c^b s_j^c(t) \bigg) . \label{eq:dcc-1-2b-comp}
\end{align}
\end{itemize}
Representative hole-mediated diagrams distinguishing connected and disconnected (but linked to the left eigenvector) contributions are shown in Figure~\ref{fig:HM transfer}. 
\textcolor{black}{These diagrams illustrate how the core-hole index enforces specific coupling topologies between $T^{(N)}$ and $T^{(N{-}1)}(t)$, distinguishing physically meaningful HM transitions from formally disconnected but linked terms.}

Incorporating the time-dependent energy difference $\Delta E_{\text{CC}}(t)$ from Eq.~\eqref{eq:deltaE}, the components contribute to the Green's function through
\begin{align}
G^R_c(t) &= -i \Theta(t) e^{-i \Delta E_{\text{CC}}(t) t} 
\left( \sum_{\mu} \tilde{O}^{\text{direct}}_{\mu}(t) 
     + \sum_{\nu} \tilde{O}^{\text{HM}}_{\nu}(t) \right).
\end{align}
The spectral contributions are then obtained via Fourier transform:
\begin{align}
A^{\text{direct}}_{\mu}(\omega) &= -\frac{1}{\pi} \, 
\text{Im}\left[\mathcal{F}\{e^{-i \Delta E_{\text{CC}}(t) t} 
\tilde{O}^{\text{direct}}_{\mu}(t)\}\right], \\
A^{\text{HM}}_{\nu}(\omega) &= -\frac{1}{\pi} \, 
\text{Im}\left[\mathcal{F}\{e^{-i \Delta E_{\text{CC}}(t) t} 
\tilde{O}^{\text{HM}}_{\nu}(t)\}\right],
\end{align}
with the total spectral function:
\begin{equation}
A(\omega) = \sum_{\mu} A^{\text{direct}}_{\mu}(\omega) 
          + \sum_{\nu} A^{\text{HM}}_{\nu}(\omega).
\end{equation}

\textcolor{black}{This decomposition provides direct access to the physical origin of satellite and shake-up features, enabling detailed interpretation of correlated excitation pathways and their dependence on the approximation level of the TD-dCC ansatz.}

%%%%%%%%%%%%%%%%%%%%%%%%%%%

\section{Computational Details}\label{sec:compu_detail}

For testing the different levels of approximations introduced in Sec.~\ref{sec:dcc_nb}, we employ the four-site single-impurity Anderson model (SIAM) as well as selected molecular systems. In the SIAM, the Hamiltonian describes a single interacting impurity orbital coupled to a set of non-interacting bath orbitals. We set the impurity (core) orbital energy to $\epsilon_c = -1.5$~a.u., assign bath on-site energies $\epsilon_{d,i} \in \{-1, 1, 2\}$~a.u., use a uniform impurity--bath hybridization of $V_i = 0.4$~a.u., and choose on-site interaction strengths of $U = 2.0$ and $3.0$~a.u. 

All molecular calculations are carried out using Dunning's \texttt{cc-pVDZ} basis~\cite{dunning1989gaussian}. For the molecular test cases, the active space is constructed from the orbitals near the highest occupied molecular orbital (HOMO), the lowest unoccupied molecular orbital (LUMO), and a deep core orbital from which the excited core-hole state is generated. \textcolor{black}{This selection ensures that the dominant quasiparticle and satellite channels are retained while keeping the active-space dimension manageable.}

Prior to all RT-EOM-CC cumulant Green's function calculations, the $N$-electron CC $T^{(N)}$ and $\Lambda^{(N)}$ amplitudes are converged to a threshold of $10^{-9}$ a.u. We only include the singles and doubles amplitudes for $T^{(N)}$ and $T^{(N-1)}$ operators for all systems in this study. Time propagation is performed by numerically integrating the ODEs~\eqref{eq:dCC-Nc} and~\eqref{eq:dCC-t} using the backward differentiation formula routine implemented in \texttt{SciPy}~\cite{virtanen2020scipy}. All real-time simulations are carried out for a total time of 900 a.u. with a fixed step size of 0.1 a.u. Throughout, a uniform spectral broadening of $\eta = 0.01$ a.u. is applied. 
\textcolor{black}{We note that the computational scaling of the TD-dCC-1 and TD-dCC-1($n$b) propagations remains identical to that of the original TD-CCSD equations, as the additional overlap and commutator terms do not introduce higher-rank tensor contractions in the SD manifold.}

We also compute the quasiparticle weight $Z$ for these spectra. The quasiparticle weight quantifies the degree of electronic correlation in the system: $Z \to 1$ corresponds to weakly correlated (nearly free) quasiparticles, while $Z \ll 1$ indicates strong many-body renormalization and substantial satellite spectral weights.
To extract $Z$, we fit the normalized spectral function $A(\omega_j)$ in the vicinity of the quasiparticle peak at $\omega_0$ to the imaginary part of a single-pole Green’s function:
\begin{equation}
A_{\text{fit}}(\omega_j; Z)
  = -\frac{1}{2\pi} \, \mathrm{Im} \!\left[ \frac{Z}{\omega_j - \omega_0 + i\eta} \right].
\label{eq:Afit}
\end{equation}
The optimal quasiparticle weight $Z$ is obtained by performing
\begin{equation}
\min \sum_{j}\| A_{\text{fit}}(\omega_j; Z) - A(\omega_j) \|_2,
\label{eq:Zfit}
\end{equation}
subject to $\sum_{j} A_{\text{fit}}(\omega_j; Z) \Delta\omega = 1$ and $Z \ge 0$. 
\textcolor{black}{This fitting approach provides a consistent measure of quasiparticle renormalization across the different TD-dCC approximation levels and serves as an additional diagnostic of many-body strength beyond visual inspection of the spectral function.}

%%%%%%%%%%%%%%%%%%%%%%%%%%

\section{Results and Discussion}\label{sec:results}

In this section, we present a comprehensive analysis of our results. We first assess the performance of the proposed approximations for the SIAM model in Section~\ref{sec:siam}. Subsequently, we extend the study to molecular systems, focusing on {H$_2$O} and {CH$_4$}. To gain further insight into the nature of the electronic excitations, we perform a cluster decomposition analysis of the excitation components as discussed in \ref{sec:comp}. Finally, we breifly discuss a quantum algorithm based on block encoding of the Hamiltonian using QSP techniques for computing the core level Green’s function and estimate the errors associated with the approximate polynomial order of the Hamiltonian.

%--------------------------

\begin{figure*}
\centering
\includegraphics[width=\linewidth]{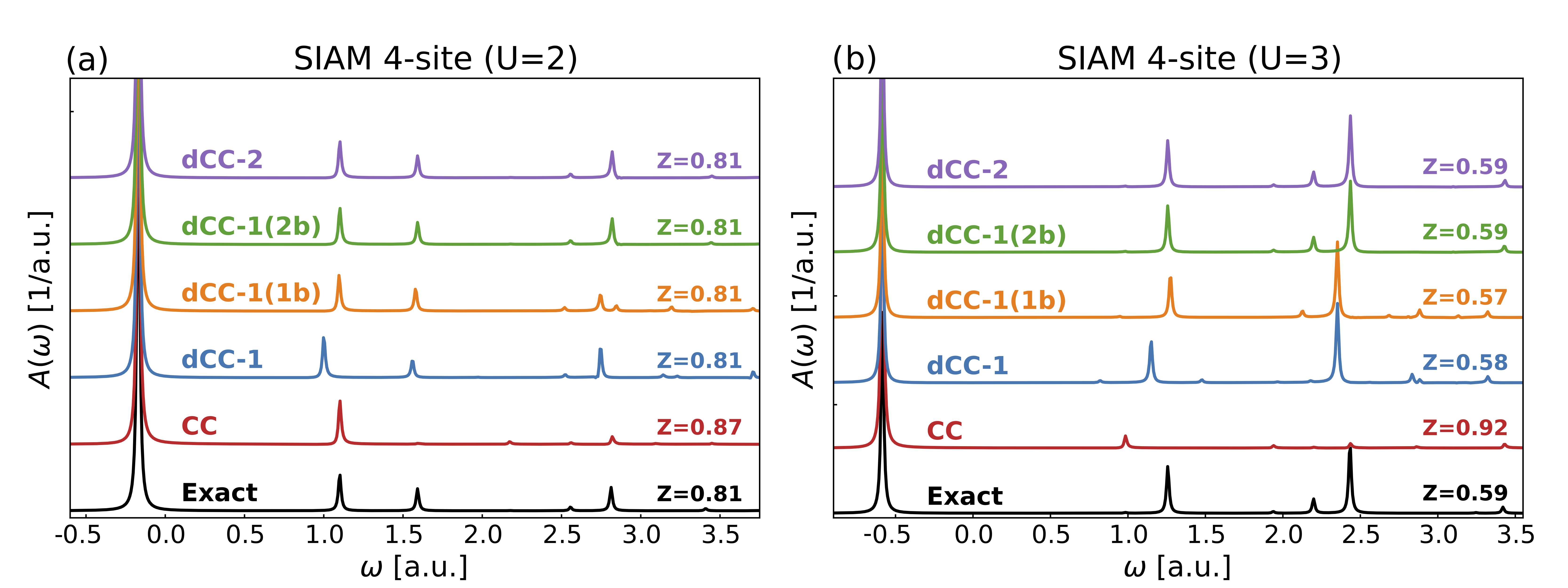}
\caption{Spectral function $A(\omega)$ of the four-site SIAM model obtained using various TD-dCC approximation schemes. (a) $U=2$ and (b) $U=3$. The corresponding quasiparticle weight $Z$ is reported for each spectrum. The main quasiparticle peak lies at 0.167~a.u. for $U=2$ and 0.586~a.u. for $U=3$. \textcolor{black}{TD-dCC-based approximations systematically recover satellite positions and spectral weights missed by TD-CC, especially in the strongly correlated $U=3$ regime.}}
\label{fig:siam_4site}
\end{figure*}

\begin{figure*}
\centering
\includegraphics[width=\linewidth]{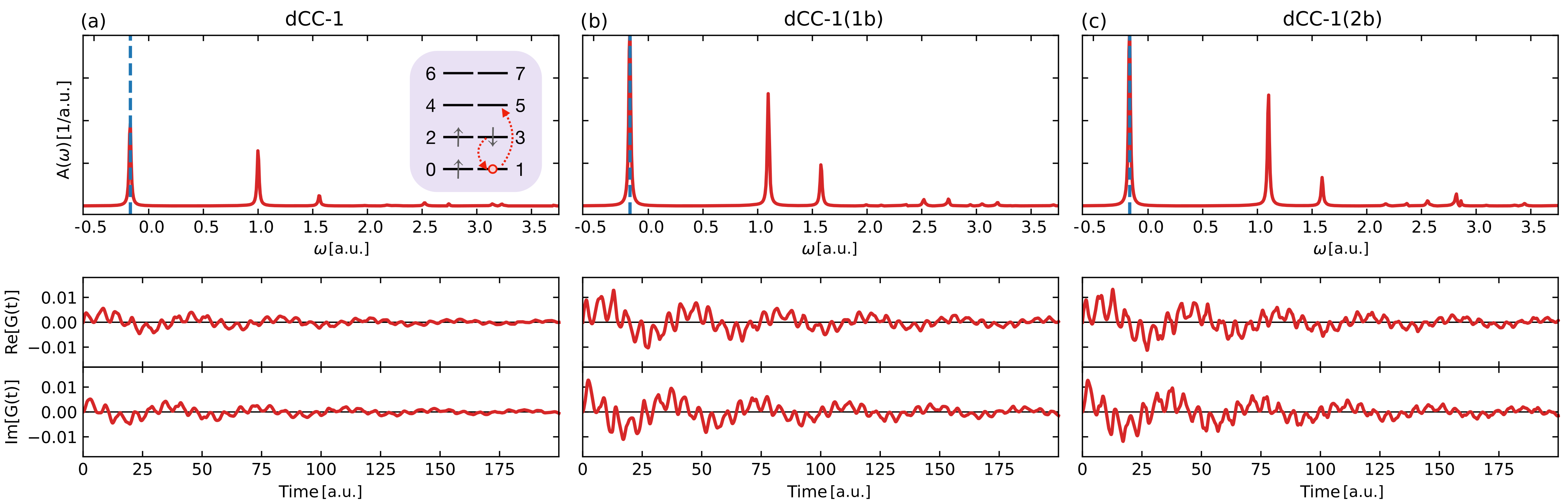}
\caption{Most significant HM transition in the four-site SIAM model ($U=2$), shown in terms of its individual spectral function and the corresponding time-dependent Green's function. \textcolor{black}{The TD-dCC hierarchy captures this correlated channel with increasing fidelity as higher-order commutator terms are included.}}
\label{fig:hm_siam}
\end{figure*}

\subsection{Four-site SIAM}\label{sec:siam}

In this section, we benchmark the accuracy of different TD-dCC approximate ans\"{a}tzes for \textcolor{black}{computing} the impurity spectral function of the four-site SIAM, and assess their ability to capture the $N$-electron correlations that are absent in the TD-CC method \eqref{eq:cc-ansatz}. 

Figure~\ref{fig:siam_4site} shows the computed impurity spectral functions $A(\omega)$ using these ans\"{a}tzes in the RT-EOM-CC cumulant Green's function framework. For comparison, we also include the exact spectral functions obtained by diagonalizing the Hamiltonian and computing the Green’s function for the impurity site. As seen, all the computed spectral functions capture the main quasiparticle peak correctly at both $U$ values. 
In the relatively weak correlation regime ($U=2$), the TD-CC ansatz \eqref{eq:cc-ansatz} (red curve) only captures one satellite feature \textcolor{black}{besides} the quasiparticle peak. In the relatively strong correlation regime ($U=3$), the TD-CC ansatz misses all the satellites between $[1.0, 3.0]$ a.u. 
\textcolor{black}{By contrast}, the TD-dCC methods provide successive corrections to the TD-CC result through incorporation of the $N$-electron ground-state correlation at different approximation levels. At both $U$ values, the lowest-level correction via the TD-dCC-1 ansatz (blue curves in Figure~\ref{fig:siam_4site}) captures the general spectral profiles, while slightly red-shifting the satellite peaks ($<0.1$ a.u.). The red-shifts are corrected by employing the single-commutator in the ansatz, as evidenced by the TD-dCC-1(1b) results (yellow curves). The higher-level corrections---TD-dCC-1(2b) and TD-dCC-2 (green and purple curves, respectively)---resolve all satellites and reproduce the exact peak positions and intensities. 

The quasiparticle weight $Z$ values from all studied approaches are also shown in Figure~\ref{fig:siam_4site}. Overall, the TD-dCC-based approximate methods capture the many-body effects more faithfully than the TD-CC approach, yielding $Z$ values close to the exact ones. In comparison, the TD-CC ansatz overestimates the $Z$ values at both $U$ values, and the discrepancy becomes larger under stronger correlation. At $U=3.0$ a.u., the TD-CC method yields $Z=0.92$, almost twice the exact value of $Z=0.59$.

Although the total spectral functions appear similar across the different TD-dCC approximations, the \textcolor{black}{underlying excitation contributions can differ substantially}. Therefore, we further perform component analyses of the computed spectral functions to explore how different parts of the TD-dCC ansatz---or distinct individual excitations---contribute to the overall spectrum. In particular, we are interested in the hole-mediated transitions, which encode how ground-state correlations transfer into the many-body dynamics of the ionized $(N-1)$-electron manifold through the core-hole channel. These HM pathways can be uniquely resolved using the analysis in Sec.~\ref{sec:comp}.

Figure~\ref{fig:hm_siam} shows the time-dependent evolution of the most significant HM single excitation (see inset in Figure~\ref{fig:hm_siam}a), revealed by the component analyses. All three approximations---TD-dCC-1, TD-dCC-1(1b), and TD-dCC-1(2b)---are capable of describing this transition. Even at the crudest level, TD-dCC-1 generates a small but non-negligible HM component. The HM contribution is slightly enhanced in the TD-dCC-1(1b) and TD-dCC-1(2b) results. Note that the TD-dCC-1(1b) and TD-dCC-1(2b) analyses yield nearly identical HM contributions, \textcolor{black}{indicating that the dominant HM process originates primarily from a singles-level coupling explicitly included in the $\Delta$ correction of TD-dCC-1(1b).}

%-----------------------------------------

\begin{figure*}
\centering
\includegraphics[width=\linewidth]{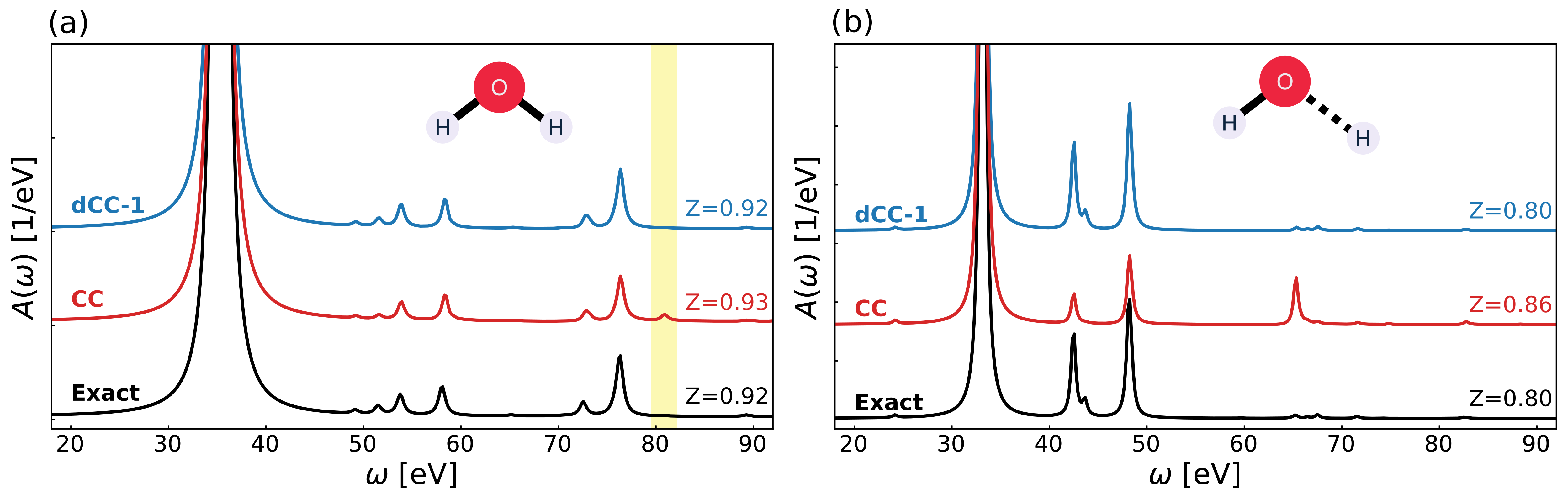}
\caption{Spectral functions of H$_2$O obtained from TD-CC, TD-dCC-1, and the exact approaches. (a) Equilibrium geometry, where the electron in the $2A_1$ orbital is ionized. The highlighted region marks a spurious extra satellite feature in the TD-CC spectrum not present in the exact or TD-dCC results. (b) Stretched geometry with one O--H bond extended to 1.5~\AA, where the electron in the $2A'$ orbital is ionized. Only TD-dCC-1 is shown since higher-level TD-dCC results are nearly indistinguishable. \textcolor{black}{Stretching the bond enhances correlation, magnifying differences between TD-CC and TD-dCC.}}
\label{fig:h2o_both}
\end{figure*}

\begin{figure}
\centering
\includegraphics[width=\linewidth]{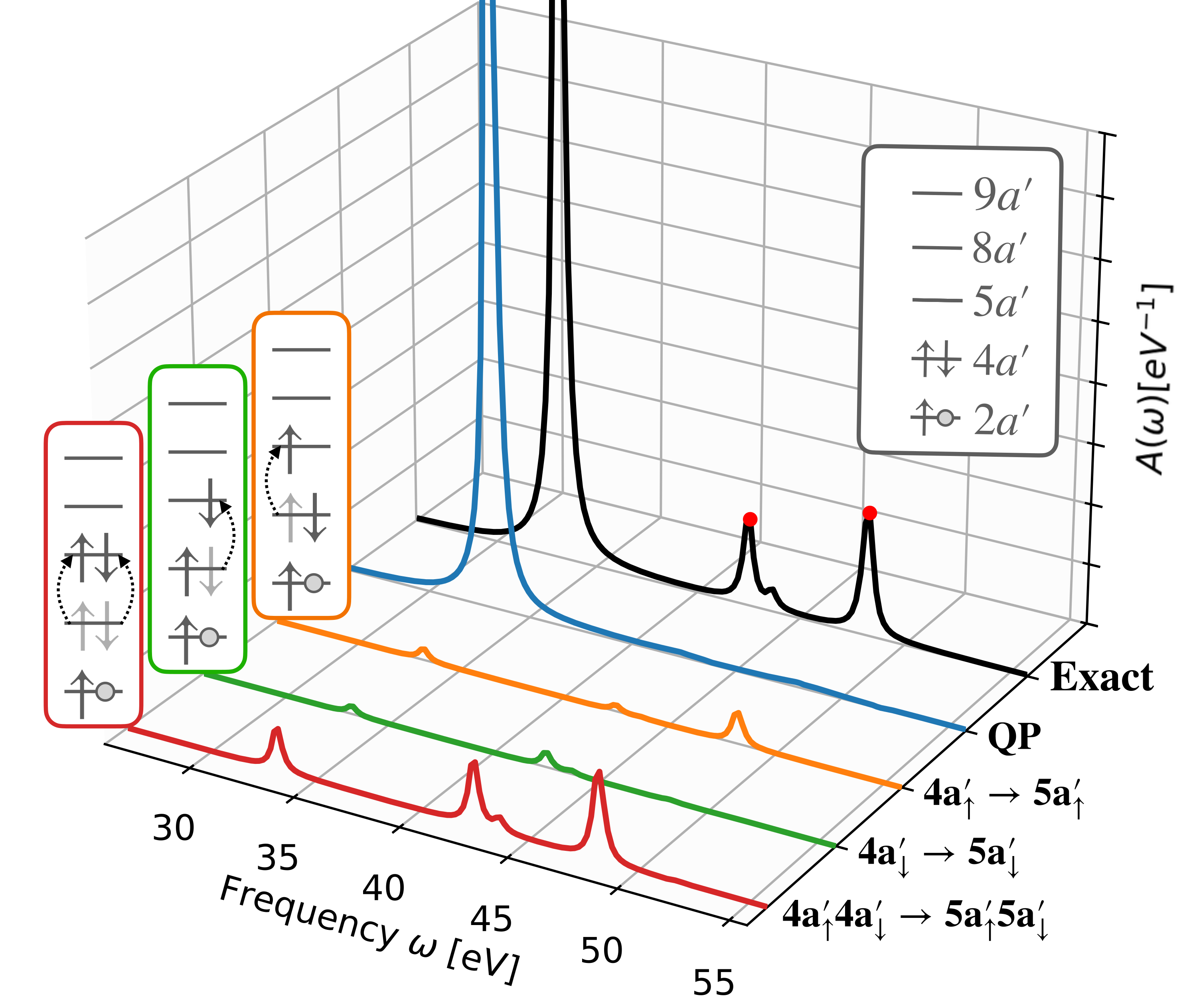}
\caption{Component analysis of the TD-dCC-1 spectral function of stretched H$_2$O, resolving the contributing single and double excitations. \textcolor{black}{The dominant satellite feature arises from the correlated double excitation $4a'_\uparrow 4a'_\downarrow \rightarrow 5a'_\uparrow 5a'_\downarrow$.}}
\label{fig:h2o_comp}
\end{figure}

\subsection{H$_2$O}

Now we switch to some simple molecular systems. For H$_2$O in the equilibrium geometry, we investigate the ionization from the oxygen $2a_1$ ($O$ 2s) orbital. Here, we limit our ionization and shake-up excitations to be within an orbital space comprising the following orbitals: $(2a_1, 3a_1, 4a_1, 5a_1, 6a_1)$, which \textcolor{black}{is chosen based on} the satellite structure analysis from the extensive CCGF study in Ref.~\citenum{peng2018green}. 

The corresponding spectral functions computed from TD-CC, TD-dCC, and exact approaches are shown in Figure~\ref{fig:h2o_both}a. Since all higher-order TD-dCC results are almost identical to the TD-dCC-1 result, only the TD-dCC-1 spectral function is shown in the plot. All three methods predict the same main quasiparticle peak at 35.32 eV, which is close to the 34.42 eV CCSD-GF result reported in Ref.~\citenum{peng2018green} with the same basis set and without active-space truncation. The quasiparticle (QP) peak and dominant spectral profile are \textcolor{black}{both} well captured at the TD-CC and TD-dCC-1 levels compared to the exact result. 

Regarding the satellite features, the TD-CC and TD-dCC-1 methods accurately capture almost all satellite peaks. For example, both approaches reproduce the largest satellite feature around 77 eV, arising from the $3a_1 \rightarrow 6a_1$ excitation (based on the peak analysis from the TD-dCC approaches). The only exception is that the TD-CC spectrum exhibits a small extra satellite feature around 80 eV (highlighted in Figure~\ref{fig:h2o_both}a), which is absent in both the exact and TD-dCC-1 spectra. In short, the water molecule in its equilibrium geometry is only weakly correlated, and the $N$-electron ground-state correlation does not significantly impact the computed spectral functions, as evidenced by the relatively large $Z$ values ($>$ 0.9).

\textcolor{black}{To explore a regime with stronger ground-state correlation}, we stretch a single O--H bond of the water molecule to $1.5~\text{\AA}$ to create a more correlated test case. For the stretched geometry, the symmetry reduces from $C_{2v}$ to $C_s$. Correspondingly, the symmetry labels in the chosen orbital space change to $2a', 4a', 5a', 8a', 9a'$.
The TD-CC, TD-dCC-1, and exact results for the stretched H$_2$O geometry are presented in Figure~\ref{fig:h2o_both}b. In this case, we observe that the TD-CC method overestimates the QP weight by approximately 0.06, whereas TD-dCC-1 accurately reproduces the exact $Z$ value. Moreover, the TD-CC ansatz fails to capture the satellite structure: for example, it misses the double peak around 42 eV and generates an artificial feature near 65 eV. By contrast, the TD-dCC approaches capture all satellite features with high accuracy. 
\textcolor{black}{These results demonstrate that TD-dCC ans\"atzes become increasingly effective when the ground state is more strongly correlated,}
which is particularly important for computing core spectra of transition-metal complexes, where more complex satellite structures arise~\cite{conradie2022xps,klevak2014charge}.

Having access to the time-dependent overlap function also allows one to compute approximate spectral decompositions. A small contribution to the satellite structure comes from the single-electron excitation $4a' \rightarrow 5a'$, while the dominant satellite feature arises from the corresponding double excitation. These transitions are presented in Figure~\ref{fig:h2o_comp}.

%--------------------------------

\begin{figure*}
\centering
\includegraphics[width=\linewidth]{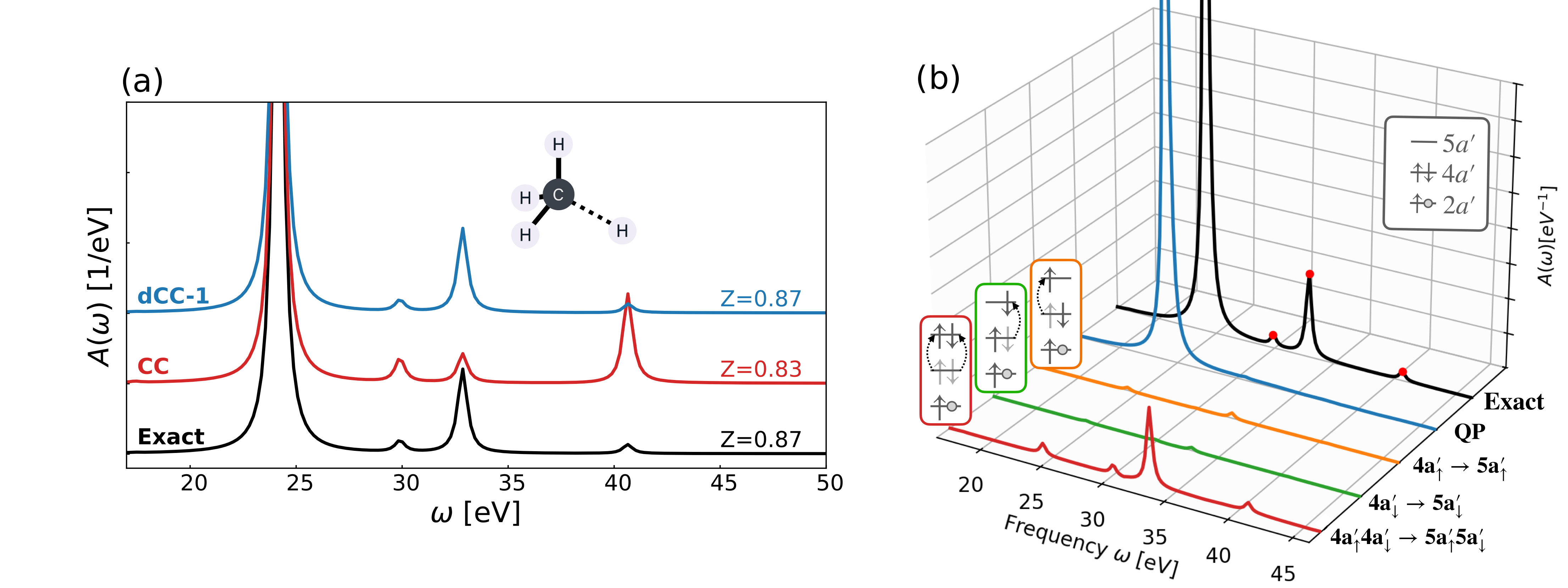}
\caption{(a) Spectral function of stretched CH$_4$ obtained from TD-CC, TD-dCC-1, and the exact reference. (b) Component analysis of the TD-dCC-1 spectral function, showing the dominant many-body channel arising from the $4a'_\uparrow 4a'_\downarrow \rightarrow 5a'_\uparrow 5a'_\downarrow$ double excitation. \textcolor{black}{TD-dCC-1 correctly captures both satellite positions and spectral weights missed by TD-CC.}}
\label{fig:ch4_comp}
\end{figure*}

\begin{figure*}
\centering
\includegraphics[width=\linewidth]{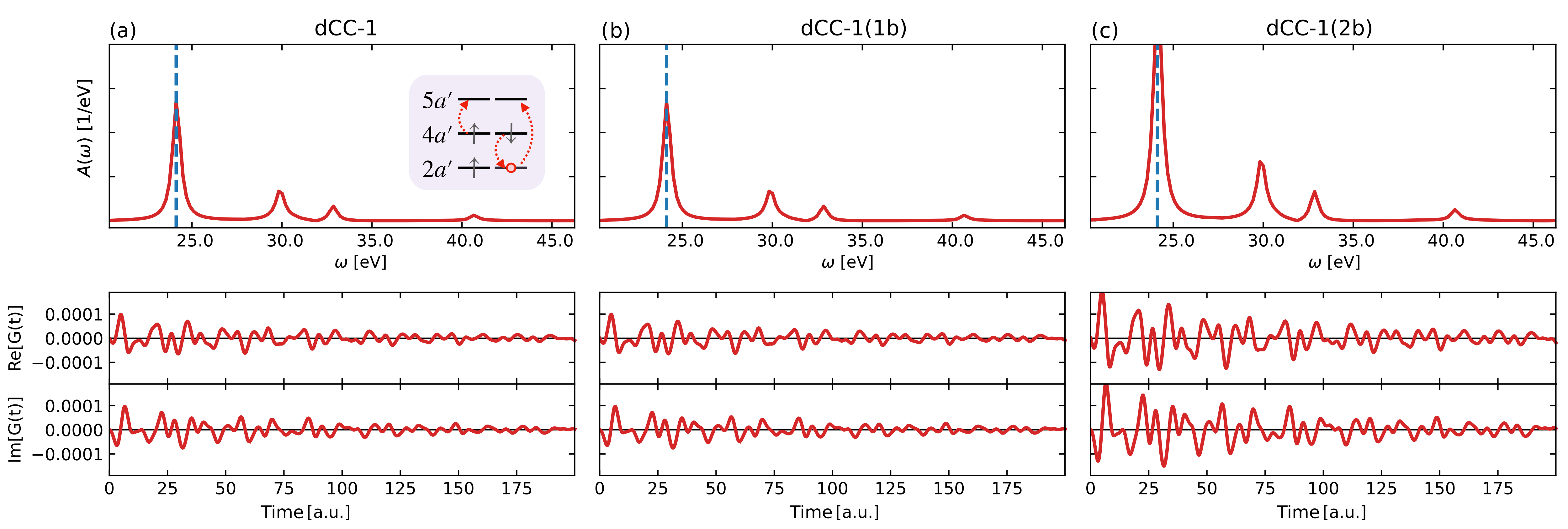}
\caption{Most significant hole-mediated (HM) transition in CH$_4$, shown as its individual spectral function and associated time-dependent Green’s function, extracted using TD-dCC-1, TD-dCC-1(1b), and TD-dCC-1(2b). \textcolor{black}{Higher-order commutator corrections enhance HM intensity by activating correlated two-electron pathways.}}
\label{fig:hm_ch4}
\end{figure*}

\subsection{CH$_4$}

Next we examine methane (CH$_4$), a prototypical closed-shell system with well-characterized spectroscopic signatures~\cite{gothe1991x,mejuto2021multi,wen2024comparing,marie2024reference}. \textcolor{black}{CH$_4$ also serves as a benchmark for time-resolved XUV and soft X-ray studies, exhibiting ultrafast Jahn--Teller distortions and rapid symmetry breaking after ionization~\cite{gonccalves2021ultrafast,ridente2023femtosecond}.} 

To probe a more strongly correlated regime, we stretch one C--H bond in the CH$_4$ equilibrium geometry to \(2~\text{\AA}\), where correlation-driven satellite and multi-quasiparticle effects become significant. This distortion lowers the molecular symmetry from \(T_d\) to \(C_s\). We focus on the ionization of the \(2a'\) orbital, which derives primarily from the carbon 2s orbital. To capture the essential physics, we employ a compact active space consisting of the \(2a'\), \(4a'\), and \(5a'\) orbitals, corresponding to the core, HOMO, and LUMO levels, respectively. At this geometry, the quasiparticle (QP) peak appears at \(24.12~\text{eV}\), compared with \(25.83~\text{eV}\) at equilibrium, both computed using the \texttt{cc-pVDZ} basis set. For reference, these values are close to previously reported EOM-CCSD results of \(23.42~\text{eV}\) at equilibrium obtained with the \texttt{aug-cc-pVQZ} basis~\cite{wen2024comparing}. 

Figure~\ref{fig:ch4_comp} shows the computed TD-CC, TD-dCC-1, and exact reference spectral functions. All other TD-dCC spectral functions are very close to TD-dCC-1 and are therefore omitted. For the most significant satellite at \(\sim33~\text{eV}\), the TD-CC approach locates the correct peak position but transfers much of the spectral weight to a satellite at \(\sim40~\text{eV}\). In contrast, TD-dCC-1 accurately reproduces both the position and spectral weight of all satellite features. In particular, TD-dCC-1 restores the correct satellite weight distribution, reproducing the exact $Z$ value. 

\textcolor{black}{The component analysis in Figure~\ref{fig:ch4_comp}b reveals that the dominant many-body channel corresponds to the double excitation \(4a'_\uparrow 4a'_\downarrow \rightarrow 5a'_\uparrow 5a'_\downarrow\), with single excitations involving the same orbitals contributing only weakly. This highlights the role of correlated two-electron motion in forming the main satellite feature.}

Although the full spectral functions obtained from the three TD-dCC-1 variants (TD-dCC-1, TD-dCC-1(1b), TD-dCC-1(2b)) appear nearly identical, the underlying time-dependent amplitudes differ. To illustrate this, we focus on the most significant hole-mediated transition in this system, which arises from a two-electron movement among five spin-orbitals: ($2a'_\downarrow$, $4a'_\uparrow$, $4a'_\downarrow$, $5a'_\uparrow$, $5a'_\downarrow$) in the $(N-1)$ sector. Here, both $4a'_\uparrow$ and $4a'_\downarrow$ are initially occupied, $2a'_\downarrow$ labels the core-hole, and $5a'_\uparrow$, $5a'_\downarrow$ are virtual orbitals. \textcolor{black}{This makes the analysis more intricate than in the SIAM case, where only one electron participates.}

Following the analysis in Sec.~\ref{sec:comp}, we find that the spectral contribution arises from a superposition of three transition channels:
\begin{itemize}
\addtolength{\itemindent}{1em} % Adds 1em horizontal space
\item[(a)] a single excitation ($4a'_\downarrow \rightarrow 2a'_\downarrow$) followed by a double excitation ($4a'_\uparrow 2a'_\downarrow \rightarrow 5a'_\uparrow 5a'_\downarrow$),
    \item[~~(b)] a double excitation ($4a'_\uparrow 4a'_\downarrow \rightarrow 5a'_\uparrow 2a'_\downarrow$) followed by a single excitation ($2a'_\downarrow \rightarrow 5a'_\downarrow$), and
    \item[~~(c)] a single excitation ($4a'_\downarrow \rightarrow 2a'_\downarrow$) followed by a single excitation ($2a'_\downarrow \rightarrow 5a'_\downarrow$), along with a simultaneous single excitation $4a'_\uparrow \rightarrow 5a'_\uparrow$. 
\end{itemize}
\textcolor{black}{Channel (c) corresponds to the disconnected but linked contributions discussed in Sec.~\ref{sec:comp}, whereas channels (a) and (b) represent genuinely connected HM pathways.}

Figure~\ref{fig:hm_ch4} shows the spectral contributions associated with these concerted transitions. The results from TD-dCC-1(1b) and TD-dCC-1 are almost identical (Figs.~\ref{fig:hm_ch4}a,b), indicating that the disconnected but linked contributions are minor for this system. In contrast, the spectral intensity of the HM transition is significantly enhanced in the TD-dCC-1(2b) result (Figure~\ref{fig:hm_ch4}c), confirming that channels (a) and (b) dominate the relevant hole-mediated processes.
\textcolor{black}{This demonstrates that even when the total spectral functions appear nearly identical, the TD-dCC hierarchy provides access to detailed underlying transition mechanisms---a key advantage for interpreting correlated dynamics in time-resolved and RIXS-type spectroscopies.}

%--------------------------------------

\subsection{Quantum computing of the core-hole Green's function}

\begin{figure}
\centering
\includegraphics[width=\linewidth]{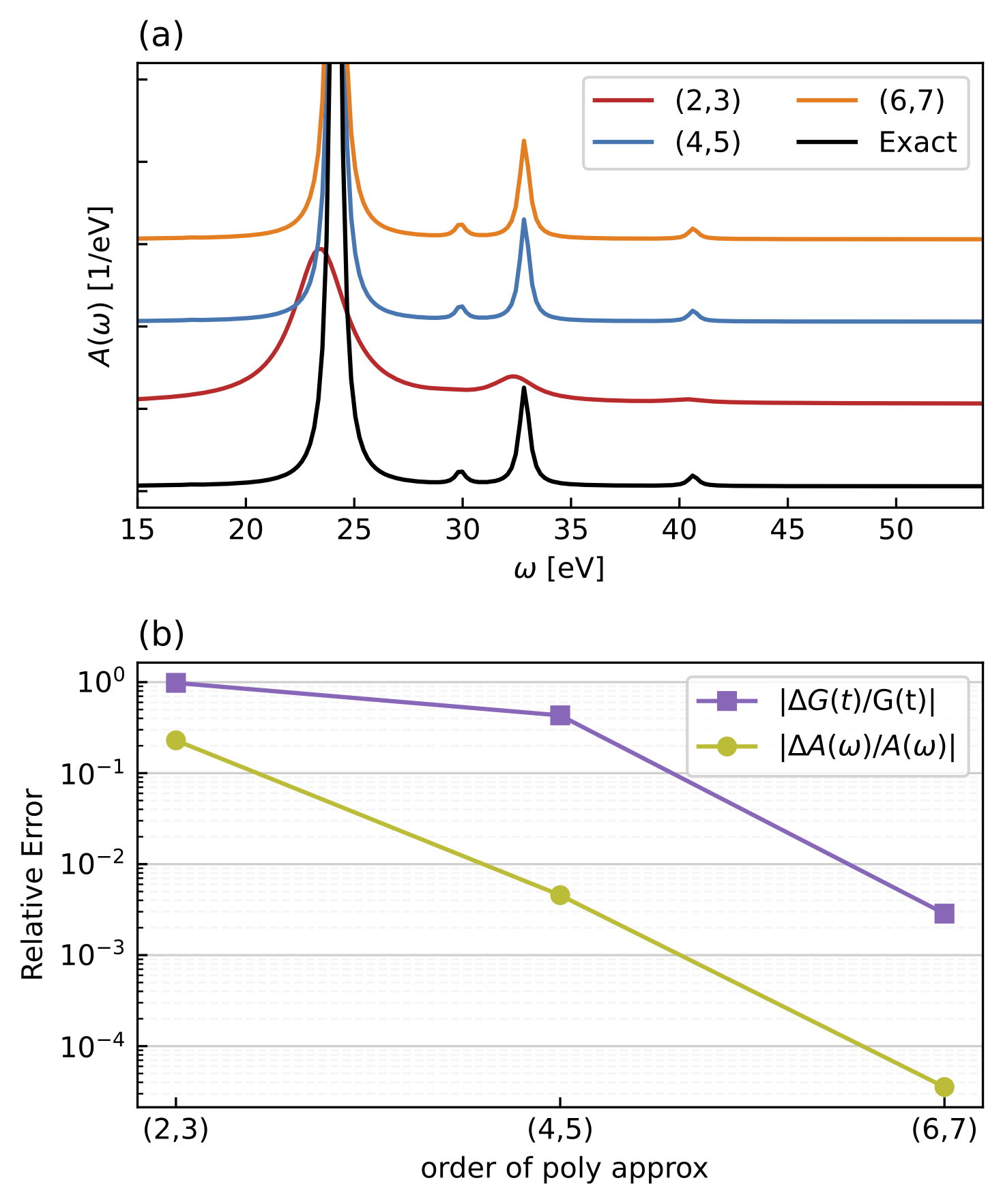}
\caption{(a) Spectral function of CH$_4$ reconstructed from the Green's function using the block-encoded Hamiltonian $H$ and the polynomial approximations to $\cos(Ht)$ and $\sin(Ht)$ for different truncation orders $(d_{\cos}, d_{\sin})$. (b) Relative mean errors in the time-dependent Green’s function and spectral function for each polynomial degree pair. \textcolor{black}{Increasing polynomial order systematically improves both $G(t)$ and $A(\omega)$, demonstrating controlled convergence of the block-encoding and QSP/QSVT approach.}}
\label{fig:qsp_poly}
\end{figure}

For multi-electron transitions that lie beyond the excitation manifold in the TD-dCC framework, the method can fail to accurately describe the associated satellite features. Including higher-order excitations can, in principle, recover these effects, but the computational cost of the required tensor contractions rises steeply---for example, coupled-cluster with triple excitations scales as $N^8$ and is impractical for systems with more than a few electrons. 
\textcolor{black}{This motivates exploring quantum algorithms, which can encode the full many-body Hamiltonian directly and potentially circumvent classical scaling limitations.}

A variety of quantum algorithms have already been proposed to evaluate many-body Green’s functions across both NISQ and fault-tolerant regimes~\cite{baker2021lanczos,kowalski2024capturing,kosugi2020construction,fomichev2025fast,kunitsa2025quantum,endo2020calculation, rizzo2022one,dhawan2024quantum,GreeneDiniz2024quantumcomputed,kanasugi2023computation}. Here, we outline a complementary perspective for computing molecular core-hole Green’s functions within a fault-tolerant framework. 

One common approach employs QSP/QSVT~\cite{Low_2017,Gily_n_2019,Martyn_2021} to evaluate the real-time Green’s function in Eq.~\eqref{eq:GF} via block encoding of the Hamiltonian. In the classical TD-dCC, the effective Hamiltonian 
\begin{align}
H_{\rm eff}^{(N-1)}(t) = e^{-T^{(N-1)}(t)} H \, e^{T^{(N-1)}(t)}
\end{align}
depends on the time-evolving cluster amplitudes $T^{(N-1)}(t)$, requiring self-consistent updates at each step. In a quantum simulation, the correlated dynamics can instead be captured by directly propagating the ionized state $a\,|\Psi^{(N)}\rangle$ under the full Hamiltonian $H$, where the unitary evolution implicitly incorporates these feedback effects.
\textcolor{black}{This removes the need for explicit time-dependent amplitude updates and shifts the complexity into the polynomial approximation used in QSP/QSVT.}
Block encoding and QSP/QSVT enable approximating 
\begin{align}
e^{iHt} = \cos(Ht) + i\sin(Ht)
\end{align}
using even/odd polynomial expansions of degrees $(d_{\cos}, d_{\sin})$~\cite{Low_2017,gilyen2019quantum,Martyn_2021,dong2022ground}. These polynomials are implemented using linear combinations of unitaries (LCU)~\cite{LCU2012,Low_2019}. All circuits in this work were generated and simulated in \textsc{PennyLane}~\cite{bergholm2018pennylane}.
\textcolor{black}{This polynomial-based simulation avoids Trotter errors and provides explicit control over the approximation error through the polynomial degree.}

Figure~\ref{fig:qsp_poly}a shows the spectral functions of CH$_4$ reconstructed from the block-encoded propagator for various truncation orders $(d_{\cos}, d_{\sin})$. The lowest-order case $(2,3)$ deviates substantially from the exact spectrum: the QP peak shifts from 24.1~eV to 23.3~eV, is overly broadened, and the satellite features are suppressed. Increasing the polynomial order to $(4,5)$ and $(6,7)$ systematically improves the fidelity, capturing both the QP peak and satellite structures. 
To quantify accuracy, Figure~\ref{fig:qsp_poly}b shows the relative mean errors in both $G(t)$ and $A(\omega)$. The lowest-order expansion exhibits large errors ($\sim$1 in $G(t)$ and 0.23 in $A(\omega)$). With $(4,5)$ the errors drop by nearly two orders of magnitude, and with $(6,7)$ they converge to $\sim 10^{-3}$ for $G(t)$ and $\sim 10^{-5}$ for $A(\omega)$.

\textcolor{black}{These results demonstrate that block encoding combined with QSP/QSVT provides a controlled, systematically improvable method for reproducing real-time Green’s functions, and therefore offers a scalable route to correlated core-level spectroscopy on future fault-tolerant quantum devices.}

%%%%%%%%%%%%%%%%%%%%%%%%%%%%%%%%%

\section{Conclusions and Outlook}\label{sec:conclusion}

We have presented a computationally efficient time-dependent double coupled-cluster (TD-dCC) framework for core-level spectroscopy that retains ground-state correlation while propagating the correlated $(N{-}1)$-electron dynamics in real time. By deriving single–exponential approximations from the first-order BCH expansion, we introduced the TD-dCC-1 hierarchy together with systematic $n$-body refinements, TD-dCC-1($n$b), that selectively incorporate many-body effects up to a chosen excitation rank, $n$. This preserves the numerical robustness and cost profile of standard RT-EOM-CC cumulant Green's function calculation while capturing essential hole-mediated correlations missing in single-CC formulations.

Benchmarking on the four-site SIAM and molecular test cases (H$_2$O and CH$_4$) shows that TD-dCC-1 and TD-dCC-1(1b/2b) deliver clear, systematic improvements over the original TD-CC ansatz: quasiparticle energies and quasiparticle weights ($Z$) are reproduced with high fidelity, and satellite structures are captured with improved accuracy when compared to exact results. In systems where ground-state correlations are significant, the TD-CC description deteriorates, whereas the TD-dCC approximations remain reliable. Therefore, the TD-dCC hierarchy provides a practical and systematically improvable framework for simulating core spectra on classical computers. 

Together with parallel developments in quantum algorithms, these real-time evolution approaches pave the way for quantitative spectroscopy of increasingly correlated molecular and materials systems. \textcolor{black}{In particular, the ability to resolve hole-mediated excitation pathways provides new interpretive insight into many-body satellite features and their dynamical origins.}

Toward a large-scale implementation, the proposed approximations maintain the formal scaling of RT-EOM-CCSD cumulant Green's function. For the SD truncation the dominant cost scales as $n^6$, and, in practice, the additional effort associated with the overlap factor is minor relative to the overall computation. Looking ahead, we will pursue a performance implementation in the Tensor Algebra for Many-body Methods (TAMM) framework~\cite{mutlu2023tamm,pathak2023real}, which has demonstrated excellent parallel scalability for large coupled-cluster calculations (e.g., Zn–porphyrin with $\sim$2{,}500 spin orbitals). We will also develop an efficient real-time numerical integrator in this computational framework~\cite{williams2023approximate,wang2022accelerating,vila2025efficient}.

On the application side, beyond photoelectron spectra, the TD-dCC approach can be extended to simulate and interpret other core spectroscopies that are sensitive to hole-mediated and multi-electron processes. In particular, RIXS and ultrafast pump–probe spectroscopies involve intermediate core-hole states whose lifetimes and correlation-driven satellite features can strongly influence the measured signals. The TD-dCC framework is well suited for simulating such time-resolved experiments, where the pump creates a non-stationary core-excited state and the subsequent probe interrogates its evolution. 

\textcolor{black}{Finally, the quantum algorithmic developments presented here suggest a promising route to scalable computation of real-time Green’s functions using block encoding and QSP/QSVT techniques. As fault-tolerant hardware matures, these approaches may enable quantitative simulation of correlated core-level dynamics far beyond the reach of classical methods.}

%%%%%%%%%%%%%%%%%%%%%%%%%%%%%%%%%

\begin{acknowledgments}
    This work is supported by the Early Career Research Program by the U.S. Department of Energy, Office of Science, under Grant No. FWP 83466. B.P. acknowledges the fruitful discussion with Dr. Karol Kowalski, Dr. Fernando Vila, and Prof. John Rehr.
\end{acknowledgments}

\bibliographystyle{achemso}
\bibliography{bibfile}

%%%%%%%%%%%%%%%%%%%%%

\appendix

\begin{figure*}
\centering
\includegraphics[width=\linewidth]{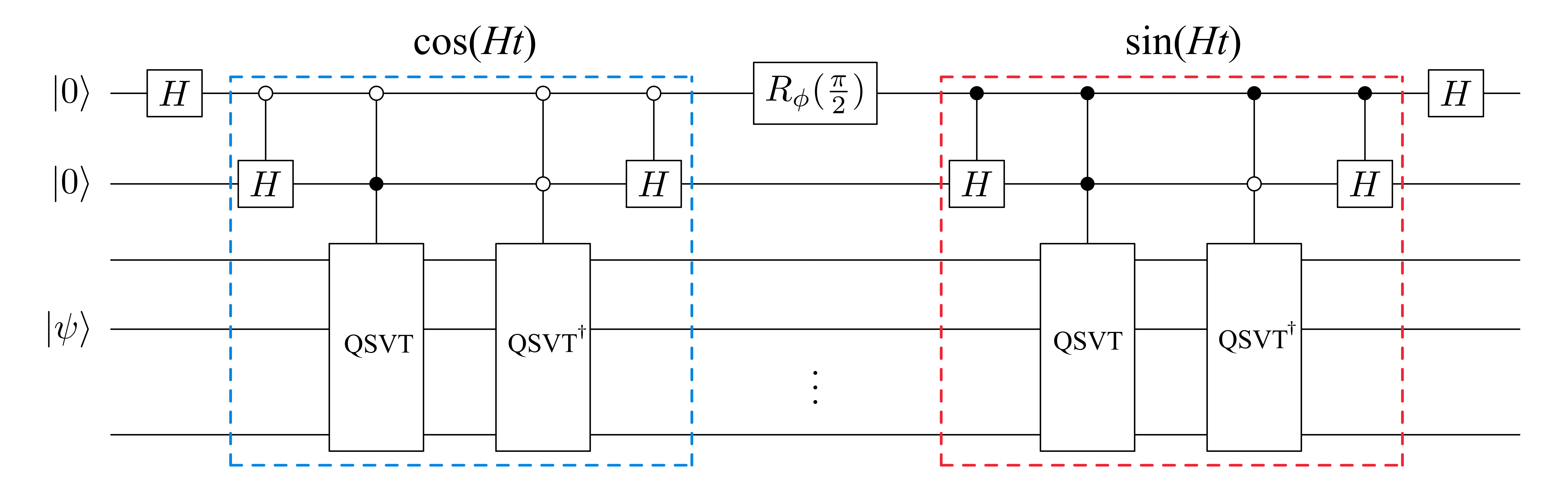}
\caption{Schematic of the QSVT-based quantum circuit implementing the cosine–sine polynomial decomposition for computing the core-hole Green's function. \textcolor{black}{The circuit uses block encoding, QSP/QSVT sequences, and LCU combination to approximate 
$e^{iHt}$ with controllable precision.}} 
\label{fig:lcu_cos}
\end{figure*}

\section{Quantum computation of Green's Function using QSVT} \label{sec:qsp}

Quantum Signal Processing (QSP)~\cite{Low_2017} and Quantum Singular Value Transformation (QSVT)~\cite{Gily_n_2019,Martyn_2021} implement polynomial transforms of block encoded operators. We adopt the \emph{embedding} block-encoding with one ancilla: for a linear operator $H$, choose $\alpha \ge \lVert H\rVert$ and realize a unitary
\begin{align}
U=\begin{bmatrix}
H/\alpha & *\\
* & *
\end{bmatrix},
\qquad \hat{H} := H/\alpha,    
\end{align}
where $\sigma(\hat{H})\subset[-1,1]$. Interleaving $U$ with a single-qubit phase sequence implements $p(\hat{H})$ for a chosen polynomial $p$ of $\hat{H}$, enabling Hamiltonian simulation and spectral transforms. Given a prepared ground state $\lvert \Psi^{(N)}\rangle$, we approximate the real-time propagator $e^{iHt}$ by QSP/QSVT polynomials for $\cos(\hat{H}t)$ and $\sin(\hat{H}t)$, i.e., $e^{i\hat{H}t}$ with a time rescaling by $\alpha$. Substituting this unitary into Eq.~\eqref{eq:GF} yields the time-domain Green's function $G(t)$, with accuracy controlled by the QSP/QSVT polynomial degree. 

We use the single-ancilla \emph{embedding} block encoding $W(\hat{H})$ satisfying 
\begin{align}
(\langle 0|\otimes I)\,W(\hat{H})\,(|0\rangle\otimes I)=\hat{H}.  
\end{align}
A $d$-depth QSP/QSVT sequence with phases $\vec\phi$ implements a bounded polynomial $p_d(\hat{H})$. For Hamiltonian simulation, choose $\vec\phi$ so that $p_d(x)\approx e^{i(\alpha t)x}$ via a truncated Chebyshev/Bessel expansion; degree $d=\Theta(\alpha t+\log(1/\varepsilon))$ achieves error $\varepsilon$~\cite{Low_2017,Gily_n_2019}. Thus one execution yields
\begin{align}
e^{iHt} \approx p_d(\hat{H}).
\end{align}
\textcolor{black}{This construction replaces Trotter decompositions with QSP-generated polynomial approximations, providing exponentially improved error scaling in the target precision at the cost of deeper but more structured circuits.}
We compile a single admissible complex polynomial $p_d(x)\approx e^{i(\alpha t)x}$, so that $\Re p_d(x)\approx \cos(\alpha t x)$ and $\Im p_d(x)\approx \sin(\alpha t x)$, avoiding extra ancillas for separating even/odd components. The time-domain Green's function then follows directly as
\begin{align}
G(t) &= \langle \Psi^{(N)} \vert\, a^\dagger\, \left(\Re p_d(\hat{H}) + i \Im p_d(\hat{H})\right)\, a \,\vert \Psi^{(N)}\rangle.
\end{align}

In our embedding setup, two degree-$d$ one-pass QSP/QSVT circuits were employed to implement $\Re p_d(\hat{H})$ and $\Im p_d(\hat{H})$ separately. The LCU technique is then used in a single pass to obtain $\Re p_d(\hat{H}) +~i \Im p_d(\hat{H})$. The circuit is shown in Figure~\ref{fig:lcu_cos}. 
\textcolor{black}{Here, a total of three ancilla qubits are used: one for the block encoding, one for the QSP/QSVT sequence, and one for the LCU combination. The ancilla count may vary under alternative block-encoding strategies, which are not investigated here.}
Note that the single-pass QSP/QSVT protocol makes exactly $d$ calls to the block-encoding unitary $W(\hat{H})$, allowing the overall circuit depth to scale linearly with the polynomial degree.

\end{document}